\documentclass[aps,prb,showpacs,twocolumn,amsmath,amssymb,superscriptaddress,letterpaper]{revtex4-1}
\usepackage{times}
\usepackage{amsfonts}
\usepackage{mathrsfs}
\usepackage{graphicx}% Include figure files
\usepackage{dcolumn}% Align table columns on decimal point
\usepackage{bm}% bold math
\usepackage{color}

\usepackage[colorlinks,bookmarks=false,citecolor=blue,linkcolor=red,urlcolor=blue]{hyperref}
\def\be{\begin{equation}}       \def\ee{\end{equation}}
\def\bea{\begin{eqnarray}}      \def\eea{\end{eqnarray}}

\begin{document}
%\begin{bibunit}

\title{ Topological Phase Transitions of Superconducting Vortex Bound States Driven by Impurities}

\author{Shengshan Qin}\email{qinshengshan@iphy.ac.cn}
\affiliation{Kavli Institute for Theoretical Sciences and CAS Center for Excellence in Topological Quantum Computation, University of Chinese Academy of Sciences, Beijing 100190, China}
\affiliation{Beijing National Research Center for Condensed Matter Physics, and Institute of Physics, Chinese Academy of Sciences, Beijing 100190, China}

\author{Chen Fang}
\affiliation{Beijing National Research Center for Condensed Matter Physics,
and Institute of Physics, Chinese Academy of Sciences, Beijing 100190, China}
\affiliation{Kavli Institute for Theoretical Sciences and CAS Center for Excellence in Topological Quantum Computation, University of Chinese Academy of Sciences, Beijing 100190, China}

\author{Fu-chun Zhang}
\affiliation{Kavli Institute for Theoretical Sciences and CAS Center for Excellence in Topological Quantum Computation, University of Chinese Academy of Sciences, Beijing 100190, China}

\author{Jiangping Hu}\email{jphu@iphy.ac.cn}
\affiliation{Beijing National Research Center for Condensed Matter Physics,
and Institute of Physics, Chinese Academy of Sciences, Beijing 100190, China}
\affiliation{Kavli Institute for Theoretical Sciences and CAS Center for Excellence in Topological Quantum Computation, University of Chinese Academy of Sciences, Beijing 100190, China}

\date{\today}

\begin{abstract}
We show that standard impurities, magnetic or nonmagnetic,   weak or strong, can cause topological phase transitions inside  the vortex cores of  a conventional s-wave superconductor. Because of the nonzero angular momentum of  Cooper pairs in the vortex cores, the vortex bound states in a two dimensional superconductor are sensitive to impurities in a way similar to  the Yu-Shiba-Rusinov bound states induced by magnetic impurities. In  three dimensional cases, the vortex bound states can be driven into topologically nontrivial phases by an impurity chain inside the vortex core. The system can host Majorana modes including the Majorana zero modes localized at the end of the vortex line and the propagating Majorana modes along the vortex line. These results suggest that the superconducting vortex can be the simplest  platform to realize Majorana modes.
\end{abstract}

%We consider a vortex line pinned by impurities in a three-dimensional (3D) conventional superconductor. We find that both nonmagnetic impurities and magnetic impurities can lead to energy crossings of the vortex bound states, akin to the condition of the Yu-Shiba-Rusinov bound states induced by magnetic impurities in a conventional superconductor. This phenomenon stems from the nonzero angular momentum of the Cooper pairs, which is contributed by the vortex. Moreover, because the gap of the vortex bound states is usually very small the energy crossings occur even if the impurity strength is rather weak. The energy crossings induced by both kinds of the impurities can drive the vortex bound states into topologically nontrivial phases. Hence Majorana modes can arise in the impurity-pinned vortices. Our study emphasize the important role of weak disorders in realizing the vortex bound Majorana modes.


\maketitle

In superconductors, vortex and impurity can induce bound states which provide the essential information of the superconductivity\cite{impurity_review,impurity_Yu,impurity_Shiba,impurity_Salkola,impurity_Hu,impurity_superfluids,impurity_Kim,impurity_Kaladzhyan, impurity_chiral_d,impurity_Guo,vortex_CdG,vortex_Hess,vortex_MacDonald,vortex_Schopohl,vortex_YBCO,vortex_SHPan,vortex_d,vortex_chiral_d}. In recent years, many studies have revealed that they may play much more fascinating roles in topological physics\cite{RMP_Sankar,RMP_Hasan,RMP_ZhangSC,RMP_Chiu,MZM_Fu,MZM_Rashba,MZM_skyrmion}. In certain conditions, they can result in the long-pursuit Majorana modes. For example, magnetic impurity chains deposited on conventional superconductors\cite{chain_Pientka,chain_Sau,chain_Brydon,chain_Heimes,chain_Bernevig} (SCs) or electrostatic impurities deposited on chiral $p$-wave SCs\cite{chain_nonmag1,chain_nonmag2} can generate Majorana zero modes (MZMs), vortices can bound MZMs in chiral $p$-wave SCs or doped superconducting topological insulators\cite{MZM_Hosur,MZM_Chiu1,MZM_Volovik,MZM_Chiu2,MZM_Fang,MZM_Xu,vortex_WTI}, and  a vortex line in superconducting Dirac semimetals can host   propagating Majorana modes\cite{MZM_DSqin,MZM_Coleman}. While the evidence of Majorana modes has been observed experimentally for some of these cases\cite{TSC_Ando,nadj2014observation,MZM_Jia,wang2018evidence,vortex_Feng,vortex_Kong,chen2019observation,wang2020evidence,MZM_flux,valentini2020flux},  it is still challenging to find the systems satisfying some of the above conditions.

In all the above proposals, the vortex and impurity are treated independently. In SCs, vortices usually are pinned by impurities  and   can be easily manipulated by artificially designed defects\cite{RMP_pinning,larkin1979pinning,pinning_Thuneberg,pinning_Klaassen,MZM_Kun}.  Thus, an  interesting  problem is the electronic physics in vortices encoded with impurities.

Here, we present a systematic study of the impurity effects  on the vortex bound states (VBSs) in $s$-wave SCs. We find that a vortex bound to a defect can be a simplest natural system to realize topological physics. Specifically,   in the diluted limit,  a single vortex in a two dimensional (2D) SC exhibits:  (i) the VBSs are prominently affected by the impurities located near the vortex core and almost immune to the impurities located out of the vortex. This phenomenon is closely related to the space localization of the VBSs; (ii)  both magnetic or nonmagnetic impurities  can  cause topological phase transitions of the VBSs by varying their strength, similar to the Yu-Shiba-Rusinov bound states induced by magnetic impurities in conventional SCs. This phenomena is caused by the fact that the Cooper pairs inside a vortex carry a nonzero angular momentum. Moreover, since the VBSs possess a minimal gap about $\Delta_0^2/2E_f$\cite{vortex_CdG} ($\Delta_0$ is the amplitude of the superconducting order and $E_f$ the Fermi energy), the phase transitions occur at a rather weak impurity strength.  Generalizing the analysis to three dimensional(3D) SCs, we find that (iii) an impurity chain in the vortex line can drive the VBSs into topologically nontrivial phases. Correspondingly, Majorana modes including the Majorana zero modes localized at the end of the vortex line and the propagating Majorana modes along the vortex line, can emerge.

\begin{figure}
\centerline{\includegraphics[width=0.45\textwidth]{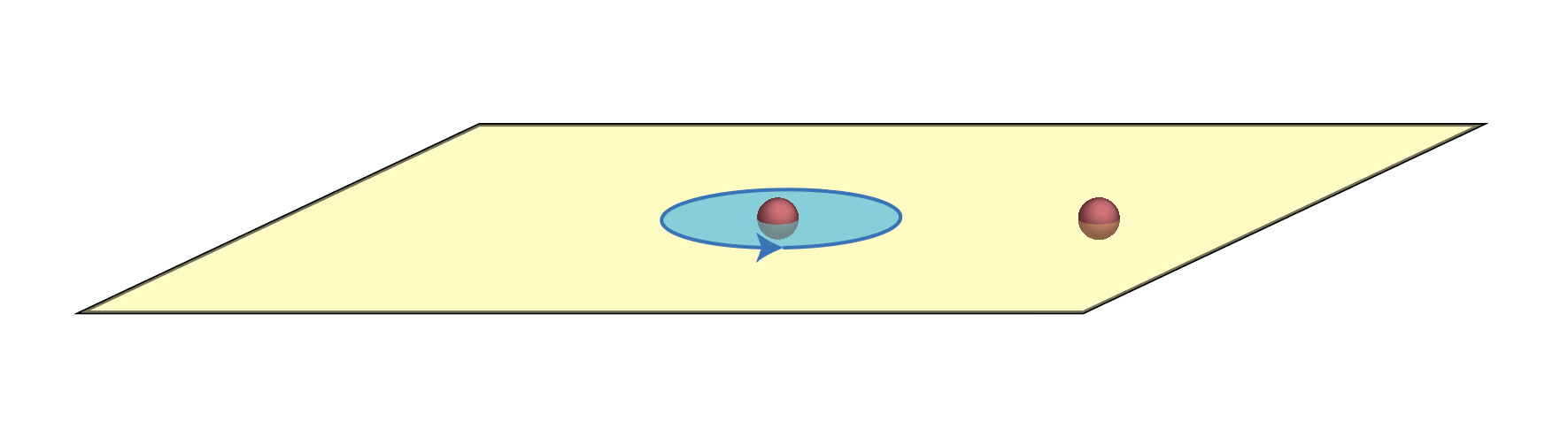}}
\caption{(color online) In type-\uppercase\expandafter{\romannumeral2} SCs, impurities can be classified into two categories according to their locations: the impurity inside the vortex affects the system like the impurity in a chiral SC, while the impurity located out of the vortex affects the system in the same way with the impurity in a conventional SC. In the figure, the vortex is labeled by the blue region and the impurity is represented by the red ball.
\label{sketch}}
\end{figure}

%We assume the vortex line spread in the superconductor straightly and the impurities distribute homogeneously in the vortex line, {\it i.e.} the impurities form a chain in the vortex line as sketched in Fig.\ref{sketch}. In such a system we find that, (i) the magnetic impurities have similar effects with nonmagnetic impurities. They both can lead to energy crossings of the VBSs, similar to the condition of the Yu-Shiba-Rusinov bound states in a conventional superconductor. This stems from the fact that, the vortex make the Cooper pairs carry a nonzero angular momentum. (ii) Because the VBSs have a minimal gap about $\Delta^2/E_f$ ($\Delta$ is the amplitude of the superconducting order and $E_f$ the Fermi energy) which usually very small, the energy crossings occur even if the impurity strength is rather weak. (iii) When spin-orbit coupling (SOC) is taken into account, the energy crossings induced by both kinds of the impurities can drive the VBSs into topologically nontrivial phases. As a result of the nontrivial topology, Majorana modes can arise in the such kind of vortices.

{\it Model Hamiltonian.} -- We begin with the 2D SCs illustrated in Fig.\ref{sketch}.  We consider  a general model with the following Hamiltonian
\begin{eqnarray}\label{H_general}
\mathcal{H} &=& \mathcal{H}_n + \mathcal{H}_{\text{sc}} + \mathcal{H}_{\text{im}},
\end{eqnarray}
where $\mathcal{H}_n$, $\mathcal{H}_{\text{sc}}$ and $\mathcal{H}_{\text{im}}$ correspond to the normal state, superconducting pairing and impurity parts of the system respectively. For a homogeneous SC, in the basis $\psi({\bf r}) = (c_{{\bf r}\uparrow}, c_{{\bf r}\downarrow}, c_{{\bf r}\downarrow}^\dag, -c_{{\bf r}\uparrow}^\dag)^T$ , the first two parts of the Hamiltonian can be specified as
\begin{eqnarray}\label{H_homogeneous}
\mathcal{H}_n + \mathcal{H}_{\text{sc}} = (\frac{{\bf k}^2}{2m} - \mu)\tau_z + \Delta({\bf r}) \tau_x,
\end{eqnarray}
where $\tau_i$ ($i = x, y, z$) are the Pauli matrices on behalf of the Nambu space. The presence of a vortex breaks the translational symmetries, and transforms the superconducting order to $\Delta({\bf r}) = \Delta_0 \tanh ( r/\xi_0)e^{i\theta}$, with $\xi_0$ the size of the vortex and $\theta$ the polar angle. The impurity part takes the form\cite{supplementary}
\begin{eqnarray}\label{H_impurity}
\mathcal{H}_{\text{im}} = S_0({\bf r}) \tau_z + {\bf S}({\bf r}) \cdot {\bf \sigma} + \mathcal{H}^{\text{im}}_{\text{soc}},
\end{eqnarray}
where ${\bf S}$ and ${\bf \sigma}$ are the spins of the impurity and the itinerant electron respectively. The first two terms in Eq.\eqref{H_impurity} are the scattering potentials of the the impurities. Specifically, $S_0$ corresponds to nonmagnetic impurity and $S_i$ the magnetic impurity. We take the impurity to be the $\delta$-function type, namely $S_\nu({\bf r}) = S_\nu \delta({\bf r-R})$ ($\nu = 0, x, y, z$) with ${\bf R}$ the location of the impurity  because the size of an impurity is usually much smaller than a vortex. Additional to the scattering potentials, a local spin-orbit coupling (SOC) term can also be induced near the impurity site since the impurity inevitably breaks the inversion symmetry locally, and it has the form
\begin{eqnarray}\label{H_impurity_SOC}
\mathcal{H}^{\text{im}}_{\text{soc}} = \alpha(r^\prime) (\hat{{\bf r}}^\prime \times {\bf k} ) \cdot {\bf \sigma} \tau_z.
\end{eqnarray}
where ${\bf r}^\prime = {\bf r - R}$, and $\hat{{\bf r}}^\prime$ is the unit vector along ${\bf r}^\prime$. We assume $\alpha(r^\prime) = \alpha_0 e^{-r^\prime/r_0}$ with $r_0 = \xi_0/2$ being the decay length of the impurity strength.

% in type-\uppercase\expandafter{\romannumeral2} SCs

We first discuss a general physical picture on the impurity effects in the type-\uppercase\expandafter{\romannumeral2} SC in an intuitive way. As described in Eq.\eqref{H_homogeneous}, the vortex breaks the homogeneity of the SC and induces a nonzero winding phase to the superconducting order around the vortex core.  We consider a single impurity.  Around the vortex core,  the impurity experiences the phase of the superconducting order around it as
\begin{eqnarray}\label{Winding_Cooper}
\gamma = \frac{1}{2\pi} \oint_l dl \Delta^{-1}({\bf r - R}) \partial_l \Delta({\bf r - R}),
\end{eqnarray}
where the integral path $l$ is the circle $| {\bf r - R} | = \xi_0$. As indicated in Fig.\ref{sketch}, it is obvious that if the impurity is located inside the vortex, $\gamma = 1$; if the impurity is located out of the vortex, $\gamma = 0$. Therefore, the impurities have completely different effects according to their locations: the impurity in the former case affects the system like the impurity in a chiral SC, while the impurity in the latter affects the system in the same way with the impurities in a conventional SC. Specifically, in a conventional SC the magnetic impurity induces Yu-Shiba-Rusinov bound states with energy $E = \frac {1 - (J \pi N/2)^2} {1 + (J \pi N/2)^2} \Delta_0$ ($J$ is the impurity strength and $N$ the density of states near the Fermi energy), and the nonmagnetic impurity only induces bound states near the edge of the superconducting gap\cite{impurity_review,impurity_Yu,impurity_Shiba}. This is consistent with the instinct that the impurity can not feel the existence of the vortex if it is located far away from the vortex.

\begin{figure}
\centerline{\includegraphics[width=0.5\textwidth]{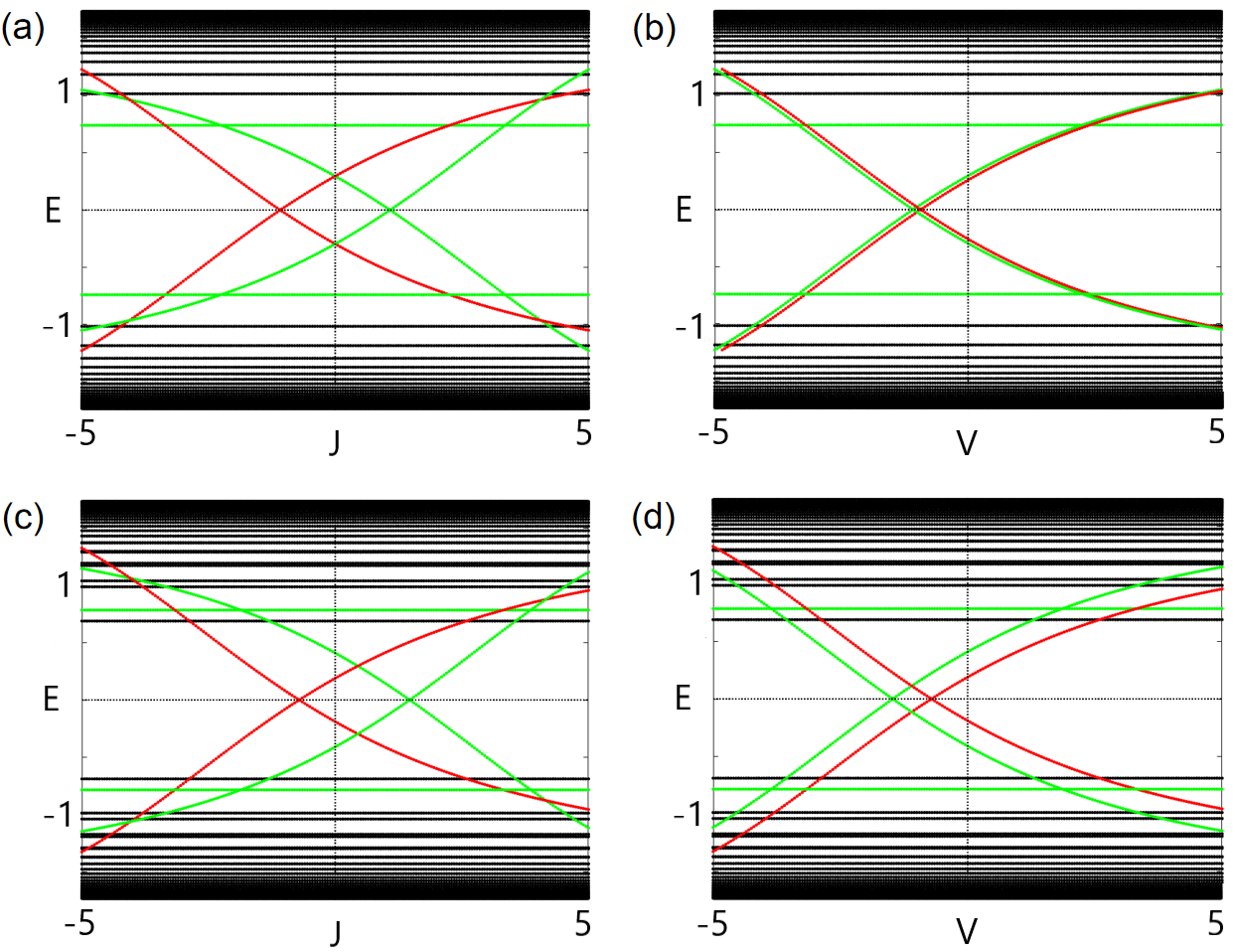}}
\caption{(color online) (a)(c) and (b)(d) show the energy spectrum of the VBSs in the condition of a magnetic impurity and a nonmagnetic impurity at the vortex core respectively. (a)(b) correspond to the limit $\alpha_0 = 0$, while $\alpha_0 = 4.0$ in (c)(d). In the figures, we label the low-energy states in the angular momentum channel $n = 0$ with red color, while the $n = \pm 1$ channel with green color. In (b), the red lines are in degenerate with the green lines and we have shifted the red lines a little to show them clearer. The other parameters are $\{ m, \mu, \Delta, \xi_0 \} = \{ 0.05, 10.0, 2.0, 2.0 \}$.
\label{fig2}}
\end{figure}

{\it Topological VBSs in 2D SCs.} -- To show the impurity effects on the VBSs when the impurity is located in the vortex, we carry out numerical simulations. For simplicity, we assume a single impurity located at the vortex core, namely ${\bf R} = (0, 0)$. Both the magnetic and nonmagnetic impurities are taken into consideration. We  also assume ${\bf S}({\bf r}) = \hat{{\bf z}} S_z \delta({\bf r})$ in the magnetic impurity case. In this condition, the rotational symmetry is preserved and the whole system can be decoupled according to the angular momentum $L_z$
\begin{eqnarray}\label{H_decouple}
\mathcal{H} &=& \oplus \mathcal{H}_{L_z}.
\end{eqnarray}
Correspondingly, the eigenstates of the system take the form $| \varphi_{L_z}({\bf r}) \rangle = e^{i( L_z - \frac{1}{2}\sigma_z + \frac{1}{2}\tau_z )\theta} \chi(r)$, with $\chi(r)$ a four-component vector. We solve the system in different angular momentum subspaces and summarize the results in Fig.\ref{fig2}. Regardless of magnetic or nonmagnetic impurities, it results in gap closures of the VBSs as the impurity strength varies, similar to the Yu-Shiba-Rusinov bound states in conventional SCs\cite{impurity_review,impurity_Yu,impurity_Shiba}. In the limit of vanishing SOC, the magnetic impurity induces two gap closures with one at $J_{c,0} = -1.08$ in the $L_z = 0$ channel and one at $J_{c,\pm1} = 1.08$ in the $L_z = \pm 1$ channel, shown in Fig.\ref{fig2}(a);  with the nonmagnetic impurity, there are also two gap closures in the $L_z = 0$ subspace and the $L_z = \pm 1$ subspace, which both occur at $V_{c,0} = V_{c,\pm 1} = -1.08$, as presented in Fig.\ref{fig2}(b). When a finite SOC is turned on, the results remain qualitatively the same except for the gap closures occurring at modified impurity strengths. As shown in Fig.\ref{fig2}(c)(d), with a finite SOC $\alpha = 4.0$ turned on, the gap closures in the magnetic impurity case are $J_{c,0} = -0.71$ and $J_{c,\pm 1} = 1.47$, while in the nonmagnetic impurity condition $V_{c,0} = -0.71$ and $V_{c,\pm 1} = -1.47$.

%Since the rotational symmetry is preserved, the eigenstates of the system in Eq.\eqref{H_general} take the form $| \varphi_{L_z}({\bf r}) \rangle = e^{i( L_z - \frac{1}{2}\sigma_z + \frac{1}{2}\tau_z )\theta} \chi(r)$, where $L_z$ is the angular momentum.

%$| \varphi_{L_z}({\bf r}) \rangle = e^{iL_z\theta} ( \chi_1(r), e^{i\theta}\chi_2(r), e^{-i\theta}\chi_3(r), \chi_4(r) )^T$

%\begin{eqnarray}\label{SOC_perturbation}
%\Delta E_{\text{SOC}} &=& -\int rd\theta dr \langle \varphi_{L_z}({\bf r}) | \frac{\alpha(r)}{r}\sigma_z\tau_z | \varphi_{L_z}({\bf r}) \rangle, \nonumber\\
%&=& -L_z \int dr \chi^\dagger(r)\sigma_z\tau_z\chi(r)\alpha(r),
%\end{eqnarray}

Fig.\ref{fig2} shows that the magnetic impurity and nonmagnetic impurity influence the VBSs in a similar way. This feature can be understood analytically.  It is  qualitatively similar to  impurity induced bound states in a chiral superconductor\cite{impurity_Kaladzhyan,impurity_chiral_d,chain_nonmag2}. We can decouple the $\delta$-function type impurity into different angular momentum channels: in the $L_z = 0$ channel a magnetic impurity is equivalent to a nonmagnetic impurity with the same scattering strength, while in the $L_z = \pm 1$ subspace a magnetic impurity is equivalent to a nonmagnetic impurity with opposite scattering strength\cite{supplementary}.

Another important feature in Fig.\ref{fig2} is that the gap closures of the VBSs occur at weak impurity strengths. This stems from the fact that, the VBSs have a minimal gap about $\Delta_0^2/2E_f$\cite{vortex_CdG} which is usually a small value compared to the bulk superconducting order $\Delta_0$.

Actually, the gap closures in Fig.\ref{fig2} mark topological phase transitions of the VBSs. In the presence of the vortex and the impurity, the translational symmetries and the time reversal symmetry of the SC are broken while the particle-hole symmetry is preserved. Therefore, the whole SC can be viewed as a quasi-0D system belonging to class-$D$ according to the Altland-Zirnbauer classification\cite{classification1,classification2}. The topological property of such a system is the $\mathcal{Z}_2$ index, which is the Pfaffian of the system and characterizes the fermion parity. In the presence of rotational symmetry, the topological property can be further considered in each of the rotational invariant subspaces\cite{rotation_Sato,rotation_CFang,rotation_RXZhang,rotation_HOTDS,rotation_fluxshell}. Since the particle-hole symmetry is antiunitary, it transforms the angular momentum $L_z$ to $-L_z$. Apparently, the particle-hole symmetry is preserved in the $L_z = 0$ subspace and broken in other subspaces. As a result, the $L_z = 0$ subsystem belongs to class-$D$ while the other subsystems belong to class-$A$ according to the Altland-Zirnbauer classification. For a 0D system belonging to class-$A$, its topological property is characterized by a $\mathcal{Z}$ index, which corresponds to the number of the states with negative energy\cite{MZM_DSqin}. However, if we consider the subsystem $\mathcal{H}_{L_z} \oplus \mathcal{H}_{-L_z}$ ($L_z \neq 0$), it regains the particle-hole symmetry. The system $\mathcal{H}_{L_z} \oplus \mathcal{H}_{-L_z}$ can be characterized by the above two different kinds of topological invariants, which are not independent and have the following relationship
\begin{eqnarray}\label{relation_topo}
\nu = e^{i\pi (N_{L_z} - N_{-L_z})/2},
\end{eqnarray}
where $N_{L_z}$ is the $\mathcal{Z}$ topological index for $\mathcal{H}_{L_z}$ and $\nu$ is the $\mathcal{Z}_2$ topological index for $\mathcal{H}_{L_z} \oplus \mathcal{H}_{-L_z}$.

In Fig.\ref{fig2}, the gap closure of the VBSs in the $L_z = 0$ channel corresponds to a change of the $\mathcal{Z}_2$ topological invariant in the $\mathcal{H}_{0}$, and the gap closure of the VBSs in the $L_z = \pm 1$ channel changes the $\mathcal{Z}_2$ ($\mathcal{Z}$) topological invariant in the $\mathcal{H}_{1} \oplus \mathcal{H}_{-1}$ ($\mathcal{H}_{1}$ and $\mathcal{H}_{-1}$). However, though the topological property of the VBSs is well-defined, it can be hardly detected since the 0D systems have no edges. This contradiction can be solved if we consider the VBSs in 3D SCs.

%The vortex transforms the superconducting order to $\Delta({\bf r}) = \Delta_0 \tanh ( r_\perp/\xi_0)e^{i\theta_\perp}$, with ${\bf r} = (x, y, z)$ and ${\bf r}_\perp = (x, y, 0)$.

%, namely ${\bf S}_\nu ({\bf r}) = {\bf S}_\nu \delta({\bf r - R})$ with ${\bf R} = (0, 0 , z)$ the locations of the impurities

{\it Topological VBSs in 3D SCs.} -- In the 3D SC, we assume a single vortex line along the $z$-direction. Obviously, the vortex line breaks the translational symmetries within the $xy$-plane and preserves the translational symmetry along the $z$-direction. Correspondingly, the whole SC can be viewed as a quasi-1D system. As pointed out in the previous study\cite{MZM_DSqin}, in the presence of the rotational symmetry the VBSs in such a system generally have four different topological states: a gapped topologically trivial state, a gapped topologically nontrivial state with MZMs at the ends of the vortex line, a nodal state having propagating Majorana modes along the vortex line, and a state where the above MZMs and propagating Majorana modes coexist in the vortex. Moreover, the above four different phases can be thoroughly characterized by the $\mathcal{Z}_2$ and $\mathcal{Z}$ topological invariants of the two 0D subsystems at the time reversal invariant momenta\cite{MZM_DSqin} (at $k_z = 0$ and $k_z = \pi$ it can viewed as two 0D subsystems of the whole quasi-1D system).

\begin{figure}
\centerline{\includegraphics[width=0.5\textwidth]{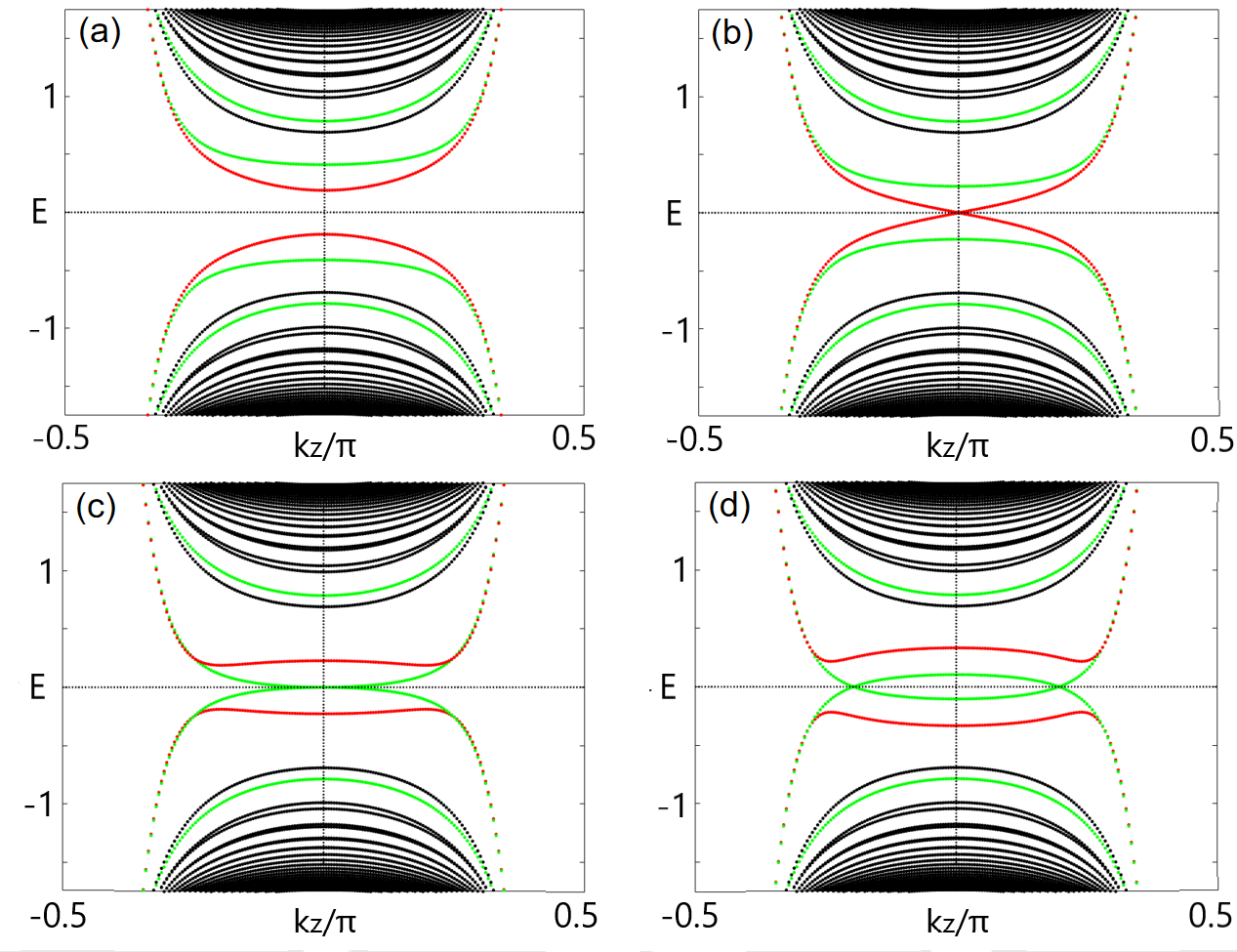}}
\caption{(color online) (a)$\sim$(d) show the energy spectrum of the VBSs in the condition of a nonmagnetic impurity chain at the core of the vortex line, as a function of $k_z$ corresponding to impurity strength $V = 0.0$, $V = -0.71$, $V = -1.47$ and $V = -1.8$ respectively. The SOC is set to be $\alpha = 4.0$. The colors in the figures and the other parameters are taken the same as that in Fig.\ref{fig2}.
\label{fig3}}
\end{figure}

In a conventional SC, the VBSs are always in the gapped topologically trivial state. However, if there are impurities which form chain-like structure in the vortex line, no matter the impurities being magnetic or nonmagnetic, the VBSs can be driven into the topologically nontrivial phases. Correspondingly, both the MZMs and the propagating Majorana modes can appear in such vortices in a conventional SC. To show this, we carry out numerical calculations in a 3D SC which can also be described by the Hamiltonian in Eq.\eqref{H_general}. We take the nonmagnetic impurity case for instance (The analysis for the magnetic impurity case is presented in the supplementary materials). We start with an ideal condition, where an impurity chain is located at the core of the vortex line and preserves the translational symmetry along the $z$-direction (the vortex line is along the $z$-direction). We take periodic boundary condition for simplicity in the calculations by transforming $k_z \rightarrow \sin k_z$ and $k_z^2 \rightarrow 2(1-\cos k_z)$. We take the same parameters with that in the 2D SC in Fig.\ref{fig2}. In this condition, the SC has a single Fermi pocket surrounding the $\Gamma$ point in the normal state leading to no low-energy VBSs at $k_z = \pi$. Thus, the topological property of the quasi-1D system is completely determined by the low-energy VBSs at $k_z = 0$. Moreover, with the above parameters the system at $k_z = 0$ is just the same as the 2D SC studied above. Straightforwardly, we come to the conclusion that the gap of the VBSs in the vortex line closes at the critical impurity strengths $V_{c,0}$ and $V_{c,\pm 1}$ in Fig.\ref{fig2}(c)(d), implying the occurrence of the topological phase transitions of the quasi-1D system. Specifically, we present the dispersion of the VBSs along $k_z$ in Fig.\ref{fig3} in which the gap closure of the VBSs at $V_{c,0} = -0.71$ and $k_z = 0$ means a topological phase transition of the VBSs in the $\mathcal{H}_{0}$ subspace.  In this case, a single MZM at each end of the vortex line appears for $V < V_{c,0}$.   Using perturbation theory as shown in supplementary, we can derive an effective Kitaev chain model\cite{Kitaev_2001} for the impurity-modified low-energy VBSs,
\begin{eqnarray}\label{Heff_zero}
h^0_{\text{eff}} = E_0(k_z) s_z + \sin k_z (s_x \text{Re} \lambda_0 - s_y \text{Im} \lambda_0),
\end{eqnarray}
where $s_i$ are the Pauli matrices representing the space spanned by the two VBSs which contribute to the topological phase transition. Similarly, the gap closure at $V_{c,\pm 1} = -1.47$ at $k_z = 0$ indicates a topological phase transition of the VBSs in the $\mathcal{H}_{L_z} \oplus \mathcal{H}_{-L_z}$ subspace. Because the gap closure occurs between VBSs with different angular momentums, it results in the propagating Majorana modes along the vortex line for $V < V_{c,\pm 1}$, as shown in Fig.\ref{fig3}(d). The above two kinds of Majorana modes can coexist for $V < V_{c,\pm 1}$ as long as the rotational symmetry is not broken.

%Therefore the low-energy effective model in this channel is
%\begin{eqnarray}\label{Heff_one}
%h_{\pm 1,\text{eff}} = (|E^{\text{min}}_1(k_z)| - |\Delta E_1|)s_z.
%\end{eqnarray}

It is necessary to consider the robustness of the above results against symmetry breaking.  There are two major cases. First,
in above analysis, if the translational symmetry along the vortex line is broken by impurities,  we can consider a superlattice structure along the vortex line which enlarges the lattice constant along the $z$-direction. The most important effect of the symmetry breaking is the folding of the Brillouin zone along $k_z$. In this case, the topological properties are robust as long as the Fermi surface topologies at $k_z = 0$ and $k_z = \pi$ stay unchanged in the the Brillouin zone folding. Second, we discuss the effect of the rotational symmetry breaking on the topological property of the VBSs.  As a result of this symmetry breaking, the VBSs in different angular momentum channels  hybridize. Therefore, the propagating Majorana modes are no longer stable, and the MZMs at the ends of the vortex line are stable in the range of $V_{c,\pm 1} < V < V_{c,0}$ ($J > J_{c,\pm 1}$ and $J < J_{c,0}$). Notice that if the rotational symmetry is absent, in the nonmagnetic impurity case the range of the topological nontrivial phase is proportional to the strength of the SOC (though the SOC considered in the paper is induced by the impurity the SOC of the bulk SC plays a similar role); while for the magnetic impurities, the VBSs are topologically nontrivial as long as the impurity strength is beyond a threshold value.

%effective model????

{\it Experimental detection.} -- The topologically nontrivial VBSs can be recognized by high-resolution scanning tunneling microscope measurements\cite{machida2019zero}. For a free vortex line in a conventional SC, its VBSs are always full-gap and topologically trivial, resulting in vanishing density of states near the zero bias in the measurements. For a vortex line with impurities if the VBSs are in topological nontrivial states, both the MZMs at the ends of the vortex line and the propagating Majorana modes along the vortex line can emerge. In the microscope scanning tunneling measurements, the former contributes to a sharp zero-bias peak near the core of the vortex, while the latter leads to nearly constant density of states around the zero bias stemming from the low-energy linear dispersion of these modes. Besides, the propagating Majorana modes can also be distinguished in the low-temperature specific heat measurements. Different from the full-gap state, these gapless modes make the specific heat scale linearly with the temperature.

{\it Discussion and conclusion.} -- In summary, we study the influences of impurities, including both the magnetic and nonmagnetic impurities, on the VBSs in the dilute limit of type-\uppercase\expandafter{\romannumeral2} SCs. We find that in 2D SCs, the VBSs are sensitive to the impurities located in the vortex and impervious to the impurities out of the vortex because of their localized nature. If the impurity is in the vortex, no matter it being magnetic or nonmagnetic, it can induce topological phase transitions of the VBSs. The nonzero angular momentum of the superconducting order contributed by the vortex plays an essential role in the topological phase transitions. Moreover, because of the small gap of the VBSs the phase transitions occur in the weak impurity region. The conclusions can be generalized to 3D SCs. It is found that if the impurities form chain-like structures in the vortex line in a 3D SC, the VBSs can be driven into the topologically nontrivial states. As a result, the MZMs localized at the ends of the vortex line and the propagating Majorana modes along the vortex line, can arise in this kind of vortices.

Our study demonstrates that the Majorana modes can appear in the vortices of a SC with disorder. For instance in SCs with dopants, the dopants inevitably introduce weak disorders which can drive the VBSs into   topologically nontrivial states.  Our study   helps to understand the recent experimental results in the iron-based SCs. In LiFeAs, it has been found that no zero energy model  is in the free vortices, while it is robustly observed  in the impurity-pinned vortices\cite{kong2020tunable}. This result can be explained by our theory if the impurities have long-range effects or the impurities form chain-like structure in the vortex line in LiFeAs.  A reversed situation occurs in (Li$_{0.84}$Fe$_{0.16}$)OHFeSe in which zero modes have been only observed in the free vortices and are absent in the vortices pinned by the dimmer defects\cite{vortex_Feng}.  The observation is believed to be a signature of the MZMs originating from the Dirac surface states\cite{vortex_Feng}. Our theory can also help to explain it.  In this condition, the MZMs originating from the surface Dirac cone hybridize with that induced by the disorders in (Li$_{0.84}$Fe$_{0.16}$)OHFeSe so that  zero energy modes  are destroyed in the defect-pinned vortices.

%In FeTe$_{0.55}$Se$_{0.45}$ and (Li$_{0.84}$Fe$_{0.16}$)OHFeSe, zero modes have been observed in the vortices, which is believed to be a signature of the MZMs originating from the Dirac surface states\cite{wang2018evidence,vortex_Feng,iron_ZJWang}. However, there are always a percentage of vortices with no zero modes in FeTe$_{0.55}$Se$_{0.45}$\cite{vortex_Kong,machida2019zero,chen2019observation}, and no robust zero modes are found in the vortices pinned by the dimmer defects in (Li$_{0.84}$Fe$_{0.16}$)OHFeSe\cite{vortex_Feng}. Considering that the substitution of Se with Te atoms in FeTe$_{0.55}$Se$_{0.45}$ and the dimmer defects in (Li$_{0.84}$Fe$_{0.16}$)OHFeSe can both introduce disorder effects, it may result in additional topological phase transitions of the VBSs. In this condition, the MZMs originating from the surface Dirac cone hybridize with that induced by the disorders. Therefore, no zero modes can be found in these vortices.

We thank C.-K. Chiu, K. Jiang, X. X. Wu and Q. Zhang for helpful discussions. This work is supported by
the Ministry of Science and Technology of China 973 program (Grant No. 2017YFA0303100),
Ministry of Science and Technology of China (Grant No. 2016YFA0302400),
National Science Foundation of China (Grant No. NSFC-11888101, Grant No. NSFC-11674370 and No. NSFC-11674278),
and Beijing Municipal Science and Technology Commission Project (Grant No. Z181100004218001),
the Strategic Priority Research Program of Chinese Academy of Sciences (Grant No. XDB28000000 and XDB33000000),
and the Information Program of the Chinese Academy of Sciences (Grant No. XXH13506-202).

%\emph{Note added}: During the preparation of this manuscript, we notice that a preprint\cite{nodal_vortex} also discuss the nodal vortex line phase in iron-based SCs when the chemical potential is near the Dirac point, which is consistent with our results.

%\begin{figure}
%\centerline{\includegraphics[width=0.45\textwidth]{fig3_new.pdf}}
%\caption{(color online)  Low energy bands in the superconducting vortex line for $Z_2^{DS}$ nontrivial and trivial DSs are presented in (a) and (b) respectively. Different colors stand for the eigenstates with different $C_{4z}^{vortex}$ eigenvalues.
%\label{vortex_pfa}}
%\end{figure}

%
%\begin{figure}
%\centerline{\includegraphics[width=0.45\textwidth]{fig4_new.pdf}}
%\caption{(color online)  Distribution of the zero modes in the $xy$-plane.
%\label{vortex_pfa}}
%\end{figure}
%

%%%%%%%%%%%%%%%%%%%%%%%%%%%%%%%%%%%%%%%%%%%%%%%%%%%%%%%%%%%%%%%%%
\bibliographystyle{apsrev4-1}
\bibliography{reference}

\begin{thebibliography}{66}%
\makeatletter
\providecommand \@ifxundefined [1]{%
 \@ifx{#1\undefined}
}%
\providecommand \@ifnum [1]{%
 \ifnum #1\expandafter \@firstoftwo
 \else \expandafter \@secondoftwo
 \fi
}%
\providecommand \@ifx [1]{%
 \ifx #1\expandafter \@firstoftwo
 \else \expandafter \@secondoftwo
 \fi
}%
\providecommand \natexlab [1]{#1}%
\providecommand \enquote  [1]{``#1''}%
\providecommand \bibnamefont  [1]{#1}%
\providecommand \bibfnamefont [1]{#1}%
\providecommand \citenamefont [1]{#1}%
\providecommand \href@noop [0]{\@secondoftwo}%
\providecommand \href [0]{\begingroup \@sanitize@url \@href}%
\providecommand \@href[1]{\@@startlink{#1}\@@href}%
\providecommand \@@href[1]{\endgroup#1\@@endlink}%
\providecommand \@sanitize@url [0]{\catcode `\\12\catcode `\$12\catcode
  `\&12\catcode `\#12\catcode `\^12\catcode `\_12\catcode `\%12\relax}%
\providecommand \@@startlink[1]{}%
\providecommand \@@endlink[0]{}%
\providecommand \url  [0]{\begingroup\@sanitize@url \@url }%
\providecommand \@url [1]{\endgroup\@href {#1}{\urlprefix }}%
\providecommand \urlprefix  [0]{URL }%
\providecommand \Eprint [0]{\href }%
\providecommand \doibase [0]{http://dx.doi.org/}%
\providecommand \selectlanguage [0]{\@gobble}%
\providecommand \bibinfo  [0]{\@secondoftwo}%
\providecommand \bibfield  [0]{\@secondoftwo}%
\providecommand \translation [1]{[#1]}%
\providecommand \BibitemOpen [0]{}%
\providecommand \bibitemStop [0]{}%
\providecommand \bibitemNoStop [0]{.\EOS\space}%
\providecommand \EOS [0]{\spacefactor3000\relax}%
\providecommand \BibitemShut  [1]{\csname bibitem#1\endcsname}%
\let\auto@bib@innerbib\@empty
%</preamble>
\bibitem [{\citenamefont {Balatsky}\ \emph {et~al.}(2006)\citenamefont
  {Balatsky}, \citenamefont {Vekhter},\ and\ \citenamefont
  {Zhu}}]{impurity_review}%
  \BibitemOpen
  \bibfield  {author} {\bibinfo {author} {\bibfnamefont {A.~V.}\ \bibnamefont
  {Balatsky}}, \bibinfo {author} {\bibfnamefont {I.}~\bibnamefont {Vekhter}}, \
  and\ \bibinfo {author} {\bibfnamefont {J.-X.}\ \bibnamefont {Zhu}},\ }\href
  {\doibase 10.1103/RevModPhys.78.373} {\bibfield  {journal} {\bibinfo
  {journal} {Rev. Mod. Phys.}\ }\textbf {\bibinfo {volume} {78}},\ \bibinfo
  {pages} {373} (\bibinfo {year} {2006})}\BibitemShut {NoStop}%
\bibitem [{\citenamefont {Yu}(1965)}]{impurity_Yu}%
  \BibitemOpen
  \bibfield  {author} {\bibinfo {author} {\bibfnamefont {L.}~\bibnamefont
  {Yu}},\ }\href {\doibase 10.7498/aps.21.75} {\bibfield  {journal} {\bibinfo
  {journal} {Acta Physica Sinica}\ }\textbf {\bibinfo {volume} {114}},\
  \bibinfo {pages} {75} (\bibinfo {year} {1965})}\BibitemShut {NoStop}%
\bibitem [{\citenamefont {Shiba}(1968)}]{impurity_Shiba}%
  \BibitemOpen
  \bibfield  {author} {\bibinfo {author} {\bibfnamefont {H.}~\bibnamefont
  {Shiba}},\ }\href {\doibase 10.1143/PTP.40.435} {\bibfield  {journal}
  {\bibinfo  {journal} {Progress of Theoretical Physics}\ }\textbf {\bibinfo
  {volume} {40}},\ \bibinfo {pages} {435} (\bibinfo {year} {1968})}\BibitemShut
  {NoStop}%
\bibitem [{\citenamefont {Salkola}\ \emph {et~al.}(1997)\citenamefont
  {Salkola}, \citenamefont {Balatsky},\ and\ \citenamefont
  {Schrieffer}}]{impurity_Salkola}%
  \BibitemOpen
  \bibfield  {author} {\bibinfo {author} {\bibfnamefont {M.~I.}\ \bibnamefont
  {Salkola}}, \bibinfo {author} {\bibfnamefont {A.~V.}\ \bibnamefont
  {Balatsky}}, \ and\ \bibinfo {author} {\bibfnamefont {J.~R.}\ \bibnamefont
  {Schrieffer}},\ }\href {\doibase 10.1103/PhysRevB.55.12648} {\bibfield
  {journal} {\bibinfo  {journal} {Phys. Rev. B}\ }\textbf {\bibinfo {volume}
  {55}},\ \bibinfo {pages} {12648} (\bibinfo {year} {1997})}\BibitemShut
  {NoStop}%
\bibitem [{\citenamefont {Tsai}\ \emph {et~al.}(2009)\citenamefont {Tsai},
  \citenamefont {Zhang}, \citenamefont {Fang},\ and\ \citenamefont
  {Hu}}]{impurity_Hu}%
  \BibitemOpen
  \bibfield  {author} {\bibinfo {author} {\bibfnamefont {W.-F.}\ \bibnamefont
  {Tsai}}, \bibinfo {author} {\bibfnamefont {Y.-Y.}\ \bibnamefont {Zhang}},
  \bibinfo {author} {\bibfnamefont {C.}~\bibnamefont {Fang}}, \ and\ \bibinfo
  {author} {\bibfnamefont {J.}~\bibnamefont {Hu}},\ }\href {\doibase
  10.1103/PhysRevB.80.064513} {\bibfield  {journal} {\bibinfo  {journal} {Phys.
  Rev. B}\ }\textbf {\bibinfo {volume} {80}},\ \bibinfo {pages} {064513}
  (\bibinfo {year} {2009})}\BibitemShut {NoStop}%
\bibitem [{\citenamefont {Hu}\ \emph {et~al.}(2013)\citenamefont {Hu},
  \citenamefont {Jiang}, \citenamefont {Pu}, \citenamefont {Chen},\ and\
  \citenamefont {Liu}}]{impurity_superfluids}%
  \BibitemOpen
  \bibfield  {author} {\bibinfo {author} {\bibfnamefont {H.}~\bibnamefont
  {Hu}}, \bibinfo {author} {\bibfnamefont {L.}~\bibnamefont {Jiang}}, \bibinfo
  {author} {\bibfnamefont {H.}~\bibnamefont {Pu}}, \bibinfo {author}
  {\bibfnamefont {Y.}~\bibnamefont {Chen}}, \ and\ \bibinfo {author}
  {\bibfnamefont {X.-J.}\ \bibnamefont {Liu}},\ }\href {\doibase
  10.1103/PhysRevLett.110.020401} {\bibfield  {journal} {\bibinfo  {journal}
  {Phys. Rev. Lett.}\ }\textbf {\bibinfo {volume} {110}},\ \bibinfo {pages}
  {020401} (\bibinfo {year} {2013})}\BibitemShut {NoStop}%
\bibitem [{\citenamefont {Kim}\ \emph {et~al.}(2015)\citenamefont {Kim},
  \citenamefont {Zhang}, \citenamefont {Rossi},\ and\ \citenamefont
  {Lutchyn}}]{impurity_Kim}%
  \BibitemOpen
  \bibfield  {author} {\bibinfo {author} {\bibfnamefont {Y.}~\bibnamefont
  {Kim}}, \bibinfo {author} {\bibfnamefont {J.}~\bibnamefont {Zhang}}, \bibinfo
  {author} {\bibfnamefont {E.}~\bibnamefont {Rossi}}, \ and\ \bibinfo {author}
  {\bibfnamefont {R.~M.}\ \bibnamefont {Lutchyn}},\ }\href {\doibase
  10.1103/PhysRevLett.114.236804} {\bibfield  {journal} {\bibinfo  {journal}
  {Phys. Rev. Lett.}\ }\textbf {\bibinfo {volume} {114}},\ \bibinfo {pages}
  {236804} (\bibinfo {year} {2015})}\BibitemShut {NoStop}%
\bibitem [{\citenamefont {Kaladzhyan}\ \emph
  {et~al.}(2016{\natexlab{a}})\citenamefont {Kaladzhyan}, \citenamefont
  {Bena},\ and\ \citenamefont {Simon}}]{impurity_Kaladzhyan}%
  \BibitemOpen
  \bibfield  {author} {\bibinfo {author} {\bibfnamefont {V.}~\bibnamefont
  {Kaladzhyan}}, \bibinfo {author} {\bibfnamefont {C.}~\bibnamefont {Bena}}, \
  and\ \bibinfo {author} {\bibfnamefont {P.}~\bibnamefont {Simon}},\ }\href
  {\doibase 10.1103/PhysRevB.93.214514} {\bibfield  {journal} {\bibinfo
  {journal} {Phys. Rev. B}\ }\textbf {\bibinfo {volume} {93}},\ \bibinfo
  {pages} {214514} (\bibinfo {year} {2016}{\natexlab{a}})}\BibitemShut
  {NoStop}%
\bibitem [{\citenamefont {Mashkoori}\ \emph {et~al.}(2017)\citenamefont
  {Mashkoori}, \citenamefont {Bjornson},\ and\ \citenamefont
  {Black-Schaffer}}]{impurity_chiral_d}%
  \BibitemOpen
  \bibfield  {author} {\bibinfo {author} {\bibfnamefont {M.}~\bibnamefont
  {Mashkoori}}, \bibinfo {author} {\bibfnamefont {K.}~\bibnamefont {Bjornson}},
  \ and\ \bibinfo {author} {\bibfnamefont {A.~M.}\ \bibnamefont
  {Black-Schaffer}},\ }\href {\doibase 10.1038/srep44107} {\bibfield  {journal}
  {\bibinfo  {journal} {Sci Rep}\ }\textbf {\bibinfo {volume} {7}},\ \bibinfo
  {pages} {44107} (\bibinfo {year} {2017})}\BibitemShut {NoStop}%
\bibitem [{\citenamefont {Guo}\ \emph {et~al.}(2017)\citenamefont {Guo},
  \citenamefont {Li},\ and\ \citenamefont {Chen}}]{impurity_Guo}%
  \BibitemOpen
  \bibfield  {author} {\bibinfo {author} {\bibfnamefont {Y.-W.}\ \bibnamefont
  {Guo}}, \bibinfo {author} {\bibfnamefont {W.}~\bibnamefont {Li}}, \ and\
  \bibinfo {author} {\bibfnamefont {Y.}~\bibnamefont {Chen}},\ }\href@noop {}
  {\bibfield  {journal} {\bibinfo  {journal} {Frontiers of Physics}\ }\textbf
  {\bibinfo {volume} {12}},\ \bibinfo {pages} {127403} (\bibinfo {year}
  {2017})}\BibitemShut {NoStop}%
\bibitem [{\citenamefont {Caroli}\ \emph {et~al.}(1964)\citenamefont {Caroli},
  \citenamefont {{De Gennes}},\ and\ \citenamefont {Matricon}}]{vortex_CdG}%
  \BibitemOpen
  \bibfield  {author} {\bibinfo {author} {\bibfnamefont {C.}~\bibnamefont
  {Caroli}}, \bibinfo {author} {\bibfnamefont {P.}~\bibnamefont {{De Gennes}}},
  \ and\ \bibinfo {author} {\bibfnamefont {J.}~\bibnamefont {Matricon}},\
  }\href {\doibase https://doi.org/10.1016/0031-9163(64)90375-0} {\bibfield
  {journal} {\bibinfo  {journal} {Physics Letters}\ }\textbf {\bibinfo {volume}
  {9}},\ \bibinfo {pages} {307 } (\bibinfo {year} {1964})}\BibitemShut
  {NoStop}%
\bibitem [{\citenamefont {Hess}\ \emph {et~al.}(1989)\citenamefont {Hess},
  \citenamefont {Robinson}, \citenamefont {Dynes}, \citenamefont {Valles},\
  and\ \citenamefont {Waszczak}}]{vortex_Hess}%
  \BibitemOpen
  \bibfield  {author} {\bibinfo {author} {\bibfnamefont {H.~F.}\ \bibnamefont
  {Hess}}, \bibinfo {author} {\bibfnamefont {R.~B.}\ \bibnamefont {Robinson}},
  \bibinfo {author} {\bibfnamefont {R.~C.}\ \bibnamefont {Dynes}}, \bibinfo
  {author} {\bibfnamefont {J.~M.}\ \bibnamefont {Valles}}, \ and\ \bibinfo
  {author} {\bibfnamefont {J.~V.}\ \bibnamefont {Waszczak}},\ }\href {\doibase
  10.1103/PhysRevLett.62.214} {\bibfield  {journal} {\bibinfo  {journal} {Phys.
  Rev. Lett.}\ }\textbf {\bibinfo {volume} {62}},\ \bibinfo {pages} {214}
  (\bibinfo {year} {1989})}\BibitemShut {NoStop}%
\bibitem [{\citenamefont {Wang}\ and\ \citenamefont
  {MacDonald}(1995)}]{vortex_MacDonald}%
  \BibitemOpen
  \bibfield  {author} {\bibinfo {author} {\bibfnamefont {Y.}~\bibnamefont
  {Wang}}\ and\ \bibinfo {author} {\bibfnamefont {A.~H.}\ \bibnamefont
  {MacDonald}},\ }\href {\doibase 10.1103/PhysRevB.52.R3876} {\bibfield
  {journal} {\bibinfo  {journal} {Phys. Rev. B}\ }\textbf {\bibinfo {volume}
  {52}},\ \bibinfo {pages} {R3876} (\bibinfo {year} {1995})}\BibitemShut
  {NoStop}%
\bibitem [{\citenamefont {Schopohl}\ and\ \citenamefont
  {Maki}(1995)}]{vortex_Schopohl}%
  \BibitemOpen
  \bibfield  {author} {\bibinfo {author} {\bibfnamefont {N.}~\bibnamefont
  {Schopohl}}\ and\ \bibinfo {author} {\bibfnamefont {K.}~\bibnamefont
  {Maki}},\ }\href {\doibase 10.1103/PhysRevB.52.490} {\bibfield  {journal}
  {\bibinfo  {journal} {Phys. Rev. B}\ }\textbf {\bibinfo {volume} {52}},\
  \bibinfo {pages} {490} (\bibinfo {year} {1995})}\BibitemShut {NoStop}%
\bibitem [{\citenamefont {Maggio-Aprile}\ \emph {et~al.}(1995)\citenamefont
  {Maggio-Aprile}, \citenamefont {Renner}, \citenamefont {Erb}, \citenamefont
  {Walker},\ and\ \citenamefont {Fischer}}]{vortex_YBCO}%
  \BibitemOpen
  \bibfield  {author} {\bibinfo {author} {\bibfnamefont {I.}~\bibnamefont
  {Maggio-Aprile}}, \bibinfo {author} {\bibfnamefont {C.}~\bibnamefont
  {Renner}}, \bibinfo {author} {\bibfnamefont {A.}~\bibnamefont {Erb}},
  \bibinfo {author} {\bibfnamefont {E.}~\bibnamefont {Walker}}, \ and\ \bibinfo
  {author} {\bibfnamefont {O.}~\bibnamefont {Fischer}},\ }\href {\doibase
  10.1103/PhysRevLett.75.2754} {\bibfield  {journal} {\bibinfo  {journal}
  {Phys. Rev. Lett.}\ }\textbf {\bibinfo {volume} {75}},\ \bibinfo {pages}
  {2754} (\bibinfo {year} {1995})}\BibitemShut {NoStop}%
\bibitem [{\citenamefont {Pan}\ \emph {et~al.}(2000)\citenamefont {Pan},
  \citenamefont {Hudson}, \citenamefont {Gupta}, \citenamefont {Ng},
  \citenamefont {Eisaki}, \citenamefont {Uchida},\ and\ \citenamefont
  {Davis}}]{vortex_SHPan}%
  \BibitemOpen
  \bibfield  {author} {\bibinfo {author} {\bibfnamefont {S.~H.}\ \bibnamefont
  {Pan}}, \bibinfo {author} {\bibfnamefont {E.~W.}\ \bibnamefont {Hudson}},
  \bibinfo {author} {\bibfnamefont {A.~K.}\ \bibnamefont {Gupta}}, \bibinfo
  {author} {\bibfnamefont {K.-W.}\ \bibnamefont {Ng}}, \bibinfo {author}
  {\bibfnamefont {H.}~\bibnamefont {Eisaki}}, \bibinfo {author} {\bibfnamefont
  {S.}~\bibnamefont {Uchida}}, \ and\ \bibinfo {author} {\bibfnamefont {J.~C.}\
  \bibnamefont {Davis}},\ }\href {\doibase 10.1103/PhysRevLett.85.1536}
  {\bibfield  {journal} {\bibinfo  {journal} {Phys. Rev. Lett.}\ }\textbf
  {\bibinfo {volume} {85}},\ \bibinfo {pages} {1536} (\bibinfo {year}
  {2000})}\BibitemShut {NoStop}%
\bibitem [{\citenamefont {{Kato, M.}}\ and\ \citenamefont {{Maki,
  K.}}(2001)}]{vortex_d}%
  \BibitemOpen
  \bibfield  {author} {\bibinfo {author} {\bibnamefont {{Kato, M.}}}\ and\
  \bibinfo {author} {\bibnamefont {{Maki, K.}}},\ }\href {\doibase
  10.1209/epl/i2001-00318-5} {\bibfield  {journal} {\bibinfo  {journal}
  {Europhys. Lett.}\ }\textbf {\bibinfo {volume} {54}},\ \bibinfo {pages} {800}
  (\bibinfo {year} {2001})}\BibitemShut {NoStop}%
\bibitem [{\citenamefont {Lee}\ and\ \citenamefont
  {Schnyder}(2016)}]{vortex_chiral_d}%
  \BibitemOpen
  \bibfield  {author} {\bibinfo {author} {\bibfnamefont {D.}~\bibnamefont
  {Lee}}\ and\ \bibinfo {author} {\bibfnamefont {A.~P.}\ \bibnamefont
  {Schnyder}},\ }\href {\doibase 10.1103/PhysRevB.93.064522} {\bibfield
  {journal} {\bibinfo  {journal} {Phys. Rev. B}\ }\textbf {\bibinfo {volume}
  {93}},\ \bibinfo {pages} {064522} (\bibinfo {year} {2016})}\BibitemShut
  {NoStop}%
\bibitem [{\citenamefont {Nayak}\ \emph {et~al.}(2008)\citenamefont {Nayak},
  \citenamefont {Simon}, \citenamefont {Stern}, \citenamefont {Freedman},\ and\
  \citenamefont {Das~Sarma}}]{RMP_Sankar}%
  \BibitemOpen
  \bibfield  {author} {\bibinfo {author} {\bibfnamefont {C.}~\bibnamefont
  {Nayak}}, \bibinfo {author} {\bibfnamefont {S.~H.}\ \bibnamefont {Simon}},
  \bibinfo {author} {\bibfnamefont {A.}~\bibnamefont {Stern}}, \bibinfo
  {author} {\bibfnamefont {M.}~\bibnamefont {Freedman}}, \ and\ \bibinfo
  {author} {\bibfnamefont {S.}~\bibnamefont {Das~Sarma}},\ }\href {\doibase
  10.1103/RevModPhys.80.1083} {\bibfield  {journal} {\bibinfo  {journal} {Rev.
  Mod. Phys.}\ }\textbf {\bibinfo {volume} {80}},\ \bibinfo {pages} {1083}
  (\bibinfo {year} {2008})}\BibitemShut {NoStop}%
\bibitem [{\citenamefont {Hasan}\ and\ \citenamefont {Kane}(2010)}]{RMP_Hasan}%
  \BibitemOpen
  \bibfield  {author} {\bibinfo {author} {\bibfnamefont {M.~Z.}\ \bibnamefont
  {Hasan}}\ and\ \bibinfo {author} {\bibfnamefont {C.~L.}\ \bibnamefont
  {Kane}},\ }\href {\doibase 10.1103/RevModPhys.82.3045} {\bibfield  {journal}
  {\bibinfo  {journal} {Rev. Mod. Phys.}\ }\textbf {\bibinfo {volume} {82}},\
  \bibinfo {pages} {3045} (\bibinfo {year} {2010})}\BibitemShut {NoStop}%
\bibitem [{\citenamefont {Qi}\ and\ \citenamefont {Zhang}(2011)}]{RMP_ZhangSC}%
  \BibitemOpen
  \bibfield  {author} {\bibinfo {author} {\bibfnamefont {X.-L.}\ \bibnamefont
  {Qi}}\ and\ \bibinfo {author} {\bibfnamefont {S.-C.}\ \bibnamefont {Zhang}},\
  }\href {\doibase 10.1103/RevModPhys.83.1057} {\bibfield  {journal} {\bibinfo
  {journal} {Rev. Mod. Phys.}\ }\textbf {\bibinfo {volume} {83}},\ \bibinfo
  {pages} {1057} (\bibinfo {year} {2011})}\BibitemShut {NoStop}%
\bibitem [{\citenamefont {Chiu}\ \emph {et~al.}(2016)\citenamefont {Chiu},
  \citenamefont {Teo}, \citenamefont {Schnyder},\ and\ \citenamefont
  {Ryu}}]{RMP_Chiu}%
  \BibitemOpen
  \bibfield  {author} {\bibinfo {author} {\bibfnamefont {C.-K.}\ \bibnamefont
  {Chiu}}, \bibinfo {author} {\bibfnamefont {J.~C.~Y.}\ \bibnamefont {Teo}},
  \bibinfo {author} {\bibfnamefont {A.~P.}\ \bibnamefont {Schnyder}}, \ and\
  \bibinfo {author} {\bibfnamefont {S.}~\bibnamefont {Ryu}},\ }\href {\doibase
  10.1103/RevModPhys.88.035005} {\bibfield  {journal} {\bibinfo  {journal}
  {Rev. Mod. Phys.}\ }\textbf {\bibinfo {volume} {88}},\ \bibinfo {pages}
  {035005} (\bibinfo {year} {2016})}\BibitemShut {NoStop}%
\bibitem [{\citenamefont {Fu}\ and\ \citenamefont {Kane}(2008)}]{MZM_Fu}%
  \BibitemOpen
  \bibfield  {author} {\bibinfo {author} {\bibfnamefont {L.}~\bibnamefont
  {Fu}}\ and\ \bibinfo {author} {\bibfnamefont {C.~L.}\ \bibnamefont {Kane}},\
  }\href {\doibase 10.1103/PhysRevLett.100.096407} {\bibfield  {journal}
  {\bibinfo  {journal} {Phys. Rev. Lett.}\ }\textbf {\bibinfo {volume} {100}},\
  \bibinfo {pages} {096407} (\bibinfo {year} {2008})}\BibitemShut {NoStop}%
\bibitem [{\citenamefont {Sau}\ \emph {et~al.}(2010)\citenamefont {Sau},
  \citenamefont {Lutchyn}, \citenamefont {Tewari},\ and\ \citenamefont
  {Das~Sarma}}]{MZM_Rashba}%
  \BibitemOpen
  \bibfield  {author} {\bibinfo {author} {\bibfnamefont {J.~D.}\ \bibnamefont
  {Sau}}, \bibinfo {author} {\bibfnamefont {R.~M.}\ \bibnamefont {Lutchyn}},
  \bibinfo {author} {\bibfnamefont {S.}~\bibnamefont {Tewari}}, \ and\ \bibinfo
  {author} {\bibfnamefont {S.}~\bibnamefont {Das~Sarma}},\ }\href {\doibase
  10.1103/PhysRevLett.104.040502} {\bibfield  {journal} {\bibinfo  {journal}
  {Phys. Rev. Lett.}\ }\textbf {\bibinfo {volume} {104}},\ \bibinfo {pages}
  {040502} (\bibinfo {year} {2010})}\BibitemShut {NoStop}%
\bibitem [{\citenamefont {Yang}\ \emph {et~al.}(2016)\citenamefont {Yang},
  \citenamefont {Stano}, \citenamefont {Klinovaja},\ and\ \citenamefont
  {Loss}}]{MZM_skyrmion}%
  \BibitemOpen
  \bibfield  {author} {\bibinfo {author} {\bibfnamefont {G.}~\bibnamefont
  {Yang}}, \bibinfo {author} {\bibfnamefont {P.}~\bibnamefont {Stano}},
  \bibinfo {author} {\bibfnamefont {J.}~\bibnamefont {Klinovaja}}, \ and\
  \bibinfo {author} {\bibfnamefont {D.}~\bibnamefont {Loss}},\ }\href {\doibase
  10.1103/PhysRevB.93.224505} {\bibfield  {journal} {\bibinfo  {journal} {Phys.
  Rev. B}\ }\textbf {\bibinfo {volume} {93}},\ \bibinfo {pages} {224505}
  (\bibinfo {year} {2016})}\BibitemShut {NoStop}%
\bibitem [{\citenamefont {Pientka}\ \emph {et~al.}(2013)\citenamefont
  {Pientka}, \citenamefont {Glazman},\ and\ \citenamefont {von
  Oppen}}]{chain_Pientka}%
  \BibitemOpen
  \bibfield  {author} {\bibinfo {author} {\bibfnamefont {F.}~\bibnamefont
  {Pientka}}, \bibinfo {author} {\bibfnamefont {L.~I.}\ \bibnamefont
  {Glazman}}, \ and\ \bibinfo {author} {\bibfnamefont {F.}~\bibnamefont {von
  Oppen}},\ }\href {\doibase 10.1103/PhysRevB.88.155420} {\bibfield  {journal}
  {\bibinfo  {journal} {Phys. Rev. B}\ }\textbf {\bibinfo {volume} {88}},\
  \bibinfo {pages} {155420} (\bibinfo {year} {2013})}\BibitemShut {NoStop}%
\bibitem [{\citenamefont {Sau}\ and\ \citenamefont {Brydon}(2015)}]{chain_Sau}%
  \BibitemOpen
  \bibfield  {author} {\bibinfo {author} {\bibfnamefont {J.~D.}\ \bibnamefont
  {Sau}}\ and\ \bibinfo {author} {\bibfnamefont {P.~M.~R.}\ \bibnamefont
  {Brydon}},\ }\href {\doibase 10.1103/PhysRevLett.115.127003} {\bibfield
  {journal} {\bibinfo  {journal} {Phys. Rev. Lett.}\ }\textbf {\bibinfo
  {volume} {115}},\ \bibinfo {pages} {127003} (\bibinfo {year}
  {2015})}\BibitemShut {NoStop}%
\bibitem [{\citenamefont {Brydon}\ \emph {et~al.}(2015)\citenamefont {Brydon},
  \citenamefont {Das~Sarma}, \citenamefont {Hui},\ and\ \citenamefont
  {Sau}}]{chain_Brydon}%
  \BibitemOpen
  \bibfield  {author} {\bibinfo {author} {\bibfnamefont {P.~M.~R.}\
  \bibnamefont {Brydon}}, \bibinfo {author} {\bibfnamefont {S.}~\bibnamefont
  {Das~Sarma}}, \bibinfo {author} {\bibfnamefont {H.-Y.}\ \bibnamefont {Hui}},
  \ and\ \bibinfo {author} {\bibfnamefont {J.~D.}\ \bibnamefont {Sau}},\ }\href
  {\doibase 10.1103/PhysRevB.91.064505} {\bibfield  {journal} {\bibinfo
  {journal} {Phys. Rev. B}\ }\textbf {\bibinfo {volume} {91}},\ \bibinfo
  {pages} {064505} (\bibinfo {year} {2015})}\BibitemShut {NoStop}%
\bibitem [{\citenamefont {Heimes}\ \emph {et~al.}(2015)\citenamefont {Heimes},
  \citenamefont {Mendler},\ and\ \citenamefont {Kotetes}}]{chain_Heimes}%
  \BibitemOpen
  \bibfield  {author} {\bibinfo {author} {\bibfnamefont {A.}~\bibnamefont
  {Heimes}}, \bibinfo {author} {\bibfnamefont {D.}~\bibnamefont {Mendler}}, \
  and\ \bibinfo {author} {\bibfnamefont {P.}~\bibnamefont {Kotetes}},\ }\href
  {\doibase 10.1088/1367-2630/17/2/023051} {\bibfield  {journal} {\bibinfo
  {journal} {New Journal of Physics}\ }\textbf {\bibinfo {volume} {17}},\
  \bibinfo {pages} {023051} (\bibinfo {year} {2015})}\BibitemShut {NoStop}%
\bibitem [{\citenamefont {Li}\ \emph {et~al.}(2016)\citenamefont {Li},
  \citenamefont {Neupert}, \citenamefont {Bernevig},\ and\ \citenamefont
  {Yazdani}}]{chain_Bernevig}%
  \BibitemOpen
  \bibfield  {author} {\bibinfo {author} {\bibfnamefont {J.}~\bibnamefont
  {Li}}, \bibinfo {author} {\bibfnamefont {T.}~\bibnamefont {Neupert}},
  \bibinfo {author} {\bibfnamefont {B.~A.}\ \bibnamefont {Bernevig}}, \ and\
  \bibinfo {author} {\bibfnamefont {A.}~\bibnamefont {Yazdani}},\ }\href
  {\doibase 10.1038/ncomms10395} {\bibfield  {journal} {\bibinfo  {journal}
  {Nat Commun}\ }\textbf {\bibinfo {volume} {7}},\ \bibinfo {pages} {10395}
  (\bibinfo {year} {2016})}\BibitemShut {NoStop}%
\bibitem [{\citenamefont {Wimmer}\ \emph {et~al.}(2010)\citenamefont {Wimmer},
  \citenamefont {Akhmerov}, \citenamefont {Medvedyeva}, \citenamefont
  {Tworzyd\l{}o},\ and\ \citenamefont {Beenakker}}]{chain_nonmag1}%
  \BibitemOpen
  \bibfield  {author} {\bibinfo {author} {\bibfnamefont {M.}~\bibnamefont
  {Wimmer}}, \bibinfo {author} {\bibfnamefont {A.~R.}\ \bibnamefont
  {Akhmerov}}, \bibinfo {author} {\bibfnamefont {M.~V.}\ \bibnamefont
  {Medvedyeva}}, \bibinfo {author} {\bibfnamefont {J.}~\bibnamefont
  {Tworzyd\l{}o}}, \ and\ \bibinfo {author} {\bibfnamefont {C.~W.~J.}\
  \bibnamefont {Beenakker}},\ }\href {\doibase 10.1103/PhysRevLett.105.046803}
  {\bibfield  {journal} {\bibinfo  {journal} {Phys. Rev. Lett.}\ }\textbf
  {\bibinfo {volume} {105}},\ \bibinfo {pages} {046803} (\bibinfo {year}
  {2010})}\BibitemShut {NoStop}%
\bibitem [{\citenamefont {Kaladzhyan}\ \emph
  {et~al.}(2016{\natexlab{b}})\citenamefont {Kaladzhyan}, \citenamefont
  {R\"ontynen}, \citenamefont {Simon},\ and\ \citenamefont
  {Ojanen}}]{chain_nonmag2}%
  \BibitemOpen
  \bibfield  {author} {\bibinfo {author} {\bibfnamefont {V.}~\bibnamefont
  {Kaladzhyan}}, \bibinfo {author} {\bibfnamefont {J.}~\bibnamefont
  {R\"ontynen}}, \bibinfo {author} {\bibfnamefont {P.}~\bibnamefont {Simon}}, \
  and\ \bibinfo {author} {\bibfnamefont {T.}~\bibnamefont {Ojanen}},\ }\href
  {\doibase 10.1103/PhysRevB.94.060505} {\bibfield  {journal} {\bibinfo
  {journal} {Phys. Rev. B}\ }\textbf {\bibinfo {volume} {94}},\ \bibinfo
  {pages} {060505} (\bibinfo {year} {2016}{\natexlab{b}})}\BibitemShut
  {NoStop}%
\bibitem [{\citenamefont {Hosur}\ \emph {et~al.}(2011)\citenamefont {Hosur},
  \citenamefont {Ghaemi}, \citenamefont {Mong},\ and\ \citenamefont
  {Vishwanath}}]{MZM_Hosur}%
  \BibitemOpen
  \bibfield  {author} {\bibinfo {author} {\bibfnamefont {P.}~\bibnamefont
  {Hosur}}, \bibinfo {author} {\bibfnamefont {P.}~\bibnamefont {Ghaemi}},
  \bibinfo {author} {\bibfnamefont {R.~S.~K.}\ \bibnamefont {Mong}}, \ and\
  \bibinfo {author} {\bibfnamefont {A.}~\bibnamefont {Vishwanath}},\ }\href
  {\doibase 10.1103/PhysRevLett.107.097001} {\bibfield  {journal} {\bibinfo
  {journal} {Phys. Rev. Lett.}\ }\textbf {\bibinfo {volume} {107}},\ \bibinfo
  {pages} {097001} (\bibinfo {year} {2011})}\BibitemShut {NoStop}%
\bibitem [{\citenamefont {Chiu}\ \emph {et~al.}(2011)\citenamefont {Chiu},
  \citenamefont {Gilbert},\ and\ \citenamefont {Hughes}}]{MZM_Chiu1}%
  \BibitemOpen
  \bibfield  {author} {\bibinfo {author} {\bibfnamefont {C.-K.}\ \bibnamefont
  {Chiu}}, \bibinfo {author} {\bibfnamefont {M.~J.}\ \bibnamefont {Gilbert}}, \
  and\ \bibinfo {author} {\bibfnamefont {T.~L.}\ \bibnamefont {Hughes}},\
  }\href {\doibase 10.1103/PhysRevB.84.144507} {\bibfield  {journal} {\bibinfo
  {journal} {Phys. Rev. B}\ }\textbf {\bibinfo {volume} {84}},\ \bibinfo
  {pages} {144507} (\bibinfo {year} {2011})}\BibitemShut {NoStop}%
\bibitem [{\citenamefont {Volovik}(2011)}]{MZM_Volovik}%
  \BibitemOpen
  \bibfield  {author} {\bibinfo {author} {\bibfnamefont {G.~E.}\ \bibnamefont
  {Volovik}},\ }\href {\doibase 10.1134/s0021364011020147} {\bibfield
  {journal} {\bibinfo  {journal} {JETP Letters}\ }\textbf {\bibinfo {volume}
  {93}},\ \bibinfo {pages} {66} (\bibinfo {year} {2011})}\BibitemShut {NoStop}%
\bibitem [{\citenamefont {Chiu}\ \emph {et~al.}(2012)\citenamefont {Chiu},
  \citenamefont {Ghaemi},\ and\ \citenamefont {Hughes}}]{MZM_Chiu2}%
  \BibitemOpen
  \bibfield  {author} {\bibinfo {author} {\bibfnamefont {C.-K.}\ \bibnamefont
  {Chiu}}, \bibinfo {author} {\bibfnamefont {P.}~\bibnamefont {Ghaemi}}, \ and\
  \bibinfo {author} {\bibfnamefont {T.~L.}\ \bibnamefont {Hughes}},\ }\href
  {\doibase 10.1103/PhysRevLett.109.237009} {\bibfield  {journal} {\bibinfo
  {journal} {Phys. Rev. Lett.}\ }\textbf {\bibinfo {volume} {109}},\ \bibinfo
  {pages} {237009} (\bibinfo {year} {2012})}\BibitemShut {NoStop}%
\bibitem [{\citenamefont {Fang}\ \emph {et~al.}(2014)\citenamefont {Fang},
  \citenamefont {Gilbert},\ and\ \citenamefont {Bernevig}}]{MZM_Fang}%
  \BibitemOpen
  \bibfield  {author} {\bibinfo {author} {\bibfnamefont {C.}~\bibnamefont
  {Fang}}, \bibinfo {author} {\bibfnamefont {M.~J.}\ \bibnamefont {Gilbert}}, \
  and\ \bibinfo {author} {\bibfnamefont {B.~A.}\ \bibnamefont {Bernevig}},\
  }\href {\doibase 10.1103/PhysRevLett.112.106401} {\bibfield  {journal}
  {\bibinfo  {journal} {Phys. Rev. Lett.}\ }\textbf {\bibinfo {volume} {112}},\
  \bibinfo {pages} {106401} (\bibinfo {year} {2014})}\BibitemShut {NoStop}%
\bibitem [{\citenamefont {Xu}\ \emph {et~al.}(2016)\citenamefont {Xu},
  \citenamefont {Lian}, \citenamefont {Tang}, \citenamefont {Qi},\ and\
  \citenamefont {Zhang}}]{MZM_Xu}%
  \BibitemOpen
  \bibfield  {author} {\bibinfo {author} {\bibfnamefont {G.}~\bibnamefont
  {Xu}}, \bibinfo {author} {\bibfnamefont {B.}~\bibnamefont {Lian}}, \bibinfo
  {author} {\bibfnamefont {P.}~\bibnamefont {Tang}}, \bibinfo {author}
  {\bibfnamefont {X.-L.}\ \bibnamefont {Qi}}, \ and\ \bibinfo {author}
  {\bibfnamefont {S.-C.}\ \bibnamefont {Zhang}},\ }\href {\doibase
  10.1103/PhysRevLett.117.047001} {\bibfield  {journal} {\bibinfo  {journal}
  {Phys. Rev. Lett.}\ }\textbf {\bibinfo {volume} {117}},\ \bibinfo {pages}
  {047001} (\bibinfo {year} {2016})}\BibitemShut {NoStop}%
\bibitem [{\citenamefont {Qin}\ \emph {et~al.}(2019{\natexlab{a}})\citenamefont
  {Qin}, \citenamefont {Hu}, \citenamefont {Wu}, \citenamefont {Dai},
  \citenamefont {Fang}, \citenamefont {Zhang},\ and\ \citenamefont
  {Hu}}]{vortex_WTI}%
  \BibitemOpen
  \bibfield  {author} {\bibinfo {author} {\bibfnamefont {S.}~\bibnamefont
  {Qin}}, \bibinfo {author} {\bibfnamefont {L.}~\bibnamefont {Hu}}, \bibinfo
  {author} {\bibfnamefont {X.}~\bibnamefont {Wu}}, \bibinfo {author}
  {\bibfnamefont {X.}~\bibnamefont {Dai}}, \bibinfo {author} {\bibfnamefont
  {C.}~\bibnamefont {Fang}}, \bibinfo {author} {\bibfnamefont {F.-C.}\
  \bibnamefont {Zhang}}, \ and\ \bibinfo {author} {\bibfnamefont
  {J.}~\bibnamefont {Hu}},\ }\href {\doibase
  https://doi.org/10.1016/j.scib.2019.07.011} {\bibfield  {journal} {\bibinfo
  {journal} {Science Bulletin}\ }\textbf {\bibinfo {volume} {64}},\ \bibinfo
  {pages} {1207 } (\bibinfo {year} {2019}{\natexlab{a}})}\BibitemShut {NoStop}%
\bibitem [{\citenamefont {Qin}\ \emph {et~al.}(2019{\natexlab{b}})\citenamefont
  {Qin}, \citenamefont {Hu}, \citenamefont {Le}, \citenamefont {Zeng},
  \citenamefont {Zhang}, \citenamefont {Fang},\ and\ \citenamefont
  {Hu}}]{MZM_DSqin}%
  \BibitemOpen
  \bibfield  {author} {\bibinfo {author} {\bibfnamefont {S.}~\bibnamefont
  {Qin}}, \bibinfo {author} {\bibfnamefont {L.}~\bibnamefont {Hu}}, \bibinfo
  {author} {\bibfnamefont {C.}~\bibnamefont {Le}}, \bibinfo {author}
  {\bibfnamefont {J.}~\bibnamefont {Zeng}}, \bibinfo {author} {\bibfnamefont
  {F.-c.}\ \bibnamefont {Zhang}}, \bibinfo {author} {\bibfnamefont
  {C.}~\bibnamefont {Fang}}, \ and\ \bibinfo {author} {\bibfnamefont
  {J.}~\bibnamefont {Hu}},\ }\href {\doibase 10.1103/PhysRevLett.123.027003}
  {\bibfield  {journal} {\bibinfo  {journal} {Phys. Rev. Lett.}\ }\textbf
  {\bibinfo {volume} {123}},\ \bibinfo {pages} {027003} (\bibinfo {year}
  {2019}{\natexlab{b}})}\BibitemShut {NoStop}%
\bibitem [{\citenamefont {K\"onig}\ and\ \citenamefont
  {Coleman}(2019)}]{MZM_Coleman}%
  \BibitemOpen
  \bibfield  {author} {\bibinfo {author} {\bibfnamefont {E.~J.}\ \bibnamefont
  {K\"onig}}\ and\ \bibinfo {author} {\bibfnamefont {P.}~\bibnamefont
  {Coleman}},\ }\href {\doibase 10.1103/PhysRevLett.122.207001} {\bibfield
  {journal} {\bibinfo  {journal} {Phys. Rev. Lett.}\ }\textbf {\bibinfo
  {volume} {122}},\ \bibinfo {pages} {207001} (\bibinfo {year}
  {2019})}\BibitemShut {NoStop}%
\bibitem [{\citenamefont {Sasaki}\ \emph {et~al.}(2011)\citenamefont {Sasaki},
  \citenamefont {Kriener}, \citenamefont {Segawa}, \citenamefont {Yada},
  \citenamefont {Tanaka}, \citenamefont {Sato},\ and\ \citenamefont
  {Ando}}]{TSC_Ando}%
  \BibitemOpen
  \bibfield  {author} {\bibinfo {author} {\bibfnamefont {S.}~\bibnamefont
  {Sasaki}}, \bibinfo {author} {\bibfnamefont {M.}~\bibnamefont {Kriener}},
  \bibinfo {author} {\bibfnamefont {K.}~\bibnamefont {Segawa}}, \bibinfo
  {author} {\bibfnamefont {K.}~\bibnamefont {Yada}}, \bibinfo {author}
  {\bibfnamefont {Y.}~\bibnamefont {Tanaka}}, \bibinfo {author} {\bibfnamefont
  {M.}~\bibnamefont {Sato}}, \ and\ \bibinfo {author} {\bibfnamefont
  {Y.}~\bibnamefont {Ando}},\ }\href {\doibase 10.1103/PhysRevLett.107.217001}
  {\bibfield  {journal} {\bibinfo  {journal} {Phys. Rev. Lett.}\ }\textbf
  {\bibinfo {volume} {107}},\ \bibinfo {pages} {217001} (\bibinfo {year}
  {2011})}\BibitemShut {NoStop}%
\bibitem [{\citenamefont {Nadj-Perge}\ \emph {et~al.}(2014)\citenamefont
  {Nadj-Perge}, \citenamefont {Drozdov}, \citenamefont {Li}, \citenamefont
  {Chen}, \citenamefont {Jeon}, \citenamefont {Seo}, \citenamefont {MacDonald},
  \citenamefont {Bernevig},\ and\ \citenamefont
  {Yazdani}}]{nadj2014observation}%
  \BibitemOpen
  \bibfield  {author} {\bibinfo {author} {\bibfnamefont {S.}~\bibnamefont
  {Nadj-Perge}}, \bibinfo {author} {\bibfnamefont {I.~K.}\ \bibnamefont
  {Drozdov}}, \bibinfo {author} {\bibfnamefont {J.}~\bibnamefont {Li}},
  \bibinfo {author} {\bibfnamefont {H.}~\bibnamefont {Chen}}, \bibinfo {author}
  {\bibfnamefont {S.}~\bibnamefont {Jeon}}, \bibinfo {author} {\bibfnamefont
  {J.}~\bibnamefont {Seo}}, \bibinfo {author} {\bibfnamefont {A.~H.}\
  \bibnamefont {MacDonald}}, \bibinfo {author} {\bibfnamefont {B.~A.}\
  \bibnamefont {Bernevig}}, \ and\ \bibinfo {author} {\bibfnamefont
  {A.}~\bibnamefont {Yazdani}},\ }\href {\doibase 10.1126/science.1259327}
  {\bibfield  {journal} {\bibinfo  {journal} {Science}\ }\textbf {\bibinfo
  {volume} {346}} (\bibinfo {year} {2014}),\
  10.1126/science.1259327}\BibitemShut {NoStop}%
\bibitem [{\citenamefont {Xu}\ \emph {et~al.}(2015)\citenamefont {Xu},
  \citenamefont {Wang}, \citenamefont {Liu}, \citenamefont {Ge}, \citenamefont
  {Yang}, \citenamefont {Liu}, \citenamefont {Xu}, \citenamefont {Guan},
  \citenamefont {Gao}, \citenamefont {Qian}, \citenamefont {Liu}, \citenamefont
  {Wang}, \citenamefont {Zhang}, \citenamefont {Xue},\ and\ \citenamefont
  {Jia}}]{MZM_Jia}%
  \BibitemOpen
  \bibfield  {author} {\bibinfo {author} {\bibfnamefont {J.-P.}\ \bibnamefont
  {Xu}}, \bibinfo {author} {\bibfnamefont {M.-X.}\ \bibnamefont {Wang}},
  \bibinfo {author} {\bibfnamefont {Z.~L.}\ \bibnamefont {Liu}}, \bibinfo
  {author} {\bibfnamefont {J.-F.}\ \bibnamefont {Ge}}, \bibinfo {author}
  {\bibfnamefont {X.}~\bibnamefont {Yang}}, \bibinfo {author} {\bibfnamefont
  {C.}~\bibnamefont {Liu}}, \bibinfo {author} {\bibfnamefont {Z.~A.}\
  \bibnamefont {Xu}}, \bibinfo {author} {\bibfnamefont {D.}~\bibnamefont
  {Guan}}, \bibinfo {author} {\bibfnamefont {C.~L.}\ \bibnamefont {Gao}},
  \bibinfo {author} {\bibfnamefont {D.}~\bibnamefont {Qian}}, \bibinfo {author}
  {\bibfnamefont {Y.}~\bibnamefont {Liu}}, \bibinfo {author} {\bibfnamefont
  {Q.-H.}\ \bibnamefont {Wang}}, \bibinfo {author} {\bibfnamefont {F.-C.}\
  \bibnamefont {Zhang}}, \bibinfo {author} {\bibfnamefont {Q.-K.}\ \bibnamefont
  {Xue}}, \ and\ \bibinfo {author} {\bibfnamefont {J.-F.}\ \bibnamefont
  {Jia}},\ }\href {\doibase 10.1103/PhysRevLett.114.017001} {\bibfield
  {journal} {\bibinfo  {journal} {Phys. Rev. Lett.}\ }\textbf {\bibinfo
  {volume} {114}},\ \bibinfo {pages} {017001} (\bibinfo {year}
  {2015})}\BibitemShut {NoStop}%
\bibitem [{\citenamefont {Wang}\ \emph {et~al.}(2018)\citenamefont {Wang},
  \citenamefont {Kong}, \citenamefont {Fan}, \citenamefont {Chen},
  \citenamefont {Zhu}, \citenamefont {Liu}, \citenamefont {Cao}, \citenamefont
  {Sun}, \citenamefont {Du}, \citenamefont {Schneeloch} \emph
  {et~al.}}]{wang2018evidence}%
  \BibitemOpen
  \bibfield  {author} {\bibinfo {author} {\bibfnamefont {D.}~\bibnamefont
  {Wang}}, \bibinfo {author} {\bibfnamefont {L.}~\bibnamefont {Kong}}, \bibinfo
  {author} {\bibfnamefont {P.}~\bibnamefont {Fan}}, \bibinfo {author}
  {\bibfnamefont {H.}~\bibnamefont {Chen}}, \bibinfo {author} {\bibfnamefont
  {S.}~\bibnamefont {Zhu}}, \bibinfo {author} {\bibfnamefont {W.}~\bibnamefont
  {Liu}}, \bibinfo {author} {\bibfnamefont {L.}~\bibnamefont {Cao}}, \bibinfo
  {author} {\bibfnamefont {Y.}~\bibnamefont {Sun}}, \bibinfo {author}
  {\bibfnamefont {S.}~\bibnamefont {Du}}, \bibinfo {author} {\bibfnamefont
  {J.}~\bibnamefont {Schneeloch}},  \emph {et~al.},\ }\href {\doibase
  10.1126/science.aao1797} {\bibfield  {journal} {\bibinfo  {journal}
  {Science}\ }\textbf {\bibinfo {volume} {362}} (\bibinfo {year} {2018}),\
  10.1126/science.aao1797}\BibitemShut {NoStop}%
\bibitem [{\citenamefont {Liu}\ \emph {et~al.}(2018)\citenamefont {Liu},
  \citenamefont {Chen}, \citenamefont {Zhang}, \citenamefont {Peng},
  \citenamefont {Yan}, \citenamefont {Wen}, \citenamefont {Lou}, \citenamefont
  {Huang}, \citenamefont {Tian}, \citenamefont {Dong}, \citenamefont {Wang},
  \citenamefont {Bao}, \citenamefont {Wang}, \citenamefont {Yin}, \citenamefont
  {Zhao},\ and\ \citenamefont {Feng}}]{vortex_Feng}%
  \BibitemOpen
  \bibfield  {author} {\bibinfo {author} {\bibfnamefont {Q.}~\bibnamefont
  {Liu}}, \bibinfo {author} {\bibfnamefont {C.}~\bibnamefont {Chen}}, \bibinfo
  {author} {\bibfnamefont {T.}~\bibnamefont {Zhang}}, \bibinfo {author}
  {\bibfnamefont {R.}~\bibnamefont {Peng}}, \bibinfo {author} {\bibfnamefont
  {Y.-J.}\ \bibnamefont {Yan}}, \bibinfo {author} {\bibfnamefont {C.-H.-P.}\
  \bibnamefont {Wen}}, \bibinfo {author} {\bibfnamefont {X.}~\bibnamefont
  {Lou}}, \bibinfo {author} {\bibfnamefont {Y.-L.}\ \bibnamefont {Huang}},
  \bibinfo {author} {\bibfnamefont {J.-P.}\ \bibnamefont {Tian}}, \bibinfo
  {author} {\bibfnamefont {X.-L.}\ \bibnamefont {Dong}}, \bibinfo {author}
  {\bibfnamefont {G.-W.}\ \bibnamefont {Wang}}, \bibinfo {author}
  {\bibfnamefont {W.-C.}\ \bibnamefont {Bao}}, \bibinfo {author} {\bibfnamefont
  {Q.-H.}\ \bibnamefont {Wang}}, \bibinfo {author} {\bibfnamefont {Z.-P.}\
  \bibnamefont {Yin}}, \bibinfo {author} {\bibfnamefont {Z.-X.}\ \bibnamefont
  {Zhao}}, \ and\ \bibinfo {author} {\bibfnamefont {D.-L.}\ \bibnamefont
  {Feng}},\ }\href {\doibase 10.1103/PhysRevX.8.041056} {\bibfield  {journal}
  {\bibinfo  {journal} {Phys. Rev. X}\ }\textbf {\bibinfo {volume} {8}},\
  \bibinfo {pages} {041056} (\bibinfo {year} {2018})}\BibitemShut {NoStop}%
\bibitem [{\citenamefont {Kong}\ \emph {et~al.}(2019)\citenamefont {Kong},
  \citenamefont {Zhu}, \citenamefont {Papaj}, \citenamefont {Chen},
  \citenamefont {Cao}, \citenamefont {Isobe}, \citenamefont {Xing},
  \citenamefont {Liu}, \citenamefont {Wang}, \citenamefont {Fan}, \citenamefont
  {Sun}, \citenamefont {Du}, \citenamefont {Schneeloch}, \citenamefont {Zhong},
  \citenamefont {Gu}, \citenamefont {Fu}, \citenamefont {Gao},\ and\
  \citenamefont {Ding}}]{vortex_Kong}%
  \BibitemOpen
  \bibfield  {author} {\bibinfo {author} {\bibfnamefont {L.}~\bibnamefont
  {Kong}}, \bibinfo {author} {\bibfnamefont {S.}~\bibnamefont {Zhu}}, \bibinfo
  {author} {\bibfnamefont {M.}~\bibnamefont {Papaj}}, \bibinfo {author}
  {\bibfnamefont {H.}~\bibnamefont {Chen}}, \bibinfo {author} {\bibfnamefont
  {L.}~\bibnamefont {Cao}}, \bibinfo {author} {\bibfnamefont {H.}~\bibnamefont
  {Isobe}}, \bibinfo {author} {\bibfnamefont {Y.}~\bibnamefont {Xing}},
  \bibinfo {author} {\bibfnamefont {W.}~\bibnamefont {Liu}}, \bibinfo {author}
  {\bibfnamefont {D.}~\bibnamefont {Wang}}, \bibinfo {author} {\bibfnamefont
  {P.}~\bibnamefont {Fan}}, \bibinfo {author} {\bibfnamefont {Y.}~\bibnamefont
  {Sun}}, \bibinfo {author} {\bibfnamefont {S.}~\bibnamefont {Du}}, \bibinfo
  {author} {\bibfnamefont {J.}~\bibnamefont {Schneeloch}}, \bibinfo {author}
  {\bibfnamefont {R.}~\bibnamefont {Zhong}}, \bibinfo {author} {\bibfnamefont
  {G.}~\bibnamefont {Gu}}, \bibinfo {author} {\bibfnamefont {L.}~\bibnamefont
  {Fu}}, \bibinfo {author} {\bibfnamefont {H.-J.}\ \bibnamefont {Gao}}, \ and\
  \bibinfo {author} {\bibfnamefont {H.}~\bibnamefont {Ding}},\ }\href {\doibase
  10.1038/s41567-019-0630-5} {\bibfield  {journal} {\bibinfo  {journal} {Nature
  Physics}\ }\textbf {\bibinfo {volume} {15}},\ \bibinfo {pages} {1181}
  (\bibinfo {year} {2019})}\BibitemShut {NoStop}%
\bibitem [{\citenamefont {Chen}\ \emph {et~al.}(2019)\citenamefont {Chen},
  \citenamefont {Chen}, \citenamefont {Duan}, \citenamefont {Zhu},
  \citenamefont {Yang},\ and\ \citenamefont {Wen}}]{chen2019observation}%
  \BibitemOpen
  \bibfield  {author} {\bibinfo {author} {\bibfnamefont {X.}~\bibnamefont
  {Chen}}, \bibinfo {author} {\bibfnamefont {M.}~\bibnamefont {Chen}}, \bibinfo
  {author} {\bibfnamefont {W.}~\bibnamefont {Duan}}, \bibinfo {author}
  {\bibfnamefont {X.}~\bibnamefont {Zhu}}, \bibinfo {author} {\bibfnamefont
  {H.}~\bibnamefont {Yang}}, \ and\ \bibinfo {author} {\bibfnamefont {H.-H.}\
  \bibnamefont {Wen}},\ }\href@noop {} {\bibfield  {journal} {\bibinfo
  {journal} {arXiv preprint arXiv:1909.01686}\ } (\bibinfo {year}
  {2019})}\BibitemShut {NoStop}%
\bibitem [{\citenamefont {Wang}\ \emph {et~al.}(2020)\citenamefont {Wang},
  \citenamefont {Rodriguez}, \citenamefont {Jiao}, \citenamefont {Howard},
  \citenamefont {Graham}, \citenamefont {Gu}, \citenamefont {Hughes},
  \citenamefont {Morr},\ and\ \citenamefont {Madhavan}}]{wang2020evidence}%
  \BibitemOpen
  \bibfield  {author} {\bibinfo {author} {\bibfnamefont {Z.}~\bibnamefont
  {Wang}}, \bibinfo {author} {\bibfnamefont {J.~O.}\ \bibnamefont {Rodriguez}},
  \bibinfo {author} {\bibfnamefont {L.}~\bibnamefont {Jiao}}, \bibinfo {author}
  {\bibfnamefont {S.}~\bibnamefont {Howard}}, \bibinfo {author} {\bibfnamefont
  {M.}~\bibnamefont {Graham}}, \bibinfo {author} {\bibfnamefont
  {G.}~\bibnamefont {Gu}}, \bibinfo {author} {\bibfnamefont {T.~L.}\
  \bibnamefont {Hughes}}, \bibinfo {author} {\bibfnamefont {D.~K.}\
  \bibnamefont {Morr}}, \ and\ \bibinfo {author} {\bibfnamefont
  {V.}~\bibnamefont {Madhavan}},\ }\href {\doibase 10.1126/science.aaw8419}
  {\bibfield  {journal} {\bibinfo  {journal} {Science}\ }\textbf {\bibinfo
  {volume} {367}} (\bibinfo {year} {2020}),\
  10.1126/science.aaw8419}\BibitemShut {NoStop}%
\bibitem [{\citenamefont {Vaitiekenas}\ \emph {et~al.}(2020)\citenamefont
  {Vaitiekenas}, \citenamefont {Winkler}, \citenamefont {van Heck},
  \citenamefont {Karzig}, \citenamefont {Deng}, \citenamefont {Flensberg},
  \citenamefont {Glazman}, \citenamefont {Nayak}, \citenamefont {Krogstrup},
  \citenamefont {Lutchyn},\ and\ \citenamefont {Marcus}}]{MZM_flux}%
  \BibitemOpen
  \bibfield  {author} {\bibinfo {author} {\bibfnamefont {S.}~\bibnamefont
  {Vaitiekenas}}, \bibinfo {author} {\bibfnamefont {G.~W.}\ \bibnamefont
  {Winkler}}, \bibinfo {author} {\bibfnamefont {B.}~\bibnamefont {van Heck}},
  \bibinfo {author} {\bibfnamefont {T.}~\bibnamefont {Karzig}}, \bibinfo
  {author} {\bibfnamefont {M.~T.}\ \bibnamefont {Deng}}, \bibinfo {author}
  {\bibfnamefont {K.}~\bibnamefont {Flensberg}}, \bibinfo {author}
  {\bibfnamefont {L.~I.}\ \bibnamefont {Glazman}}, \bibinfo {author}
  {\bibfnamefont {C.}~\bibnamefont {Nayak}}, \bibinfo {author} {\bibfnamefont
  {P.}~\bibnamefont {Krogstrup}}, \bibinfo {author} {\bibfnamefont {R.~M.}\
  \bibnamefont {Lutchyn}}, \ and\ \bibinfo {author} {\bibfnamefont {C.~M.}\
  \bibnamefont {Marcus}},\ }\href {\doibase 10.1126/science.aav3392} {\bibfield
   {journal} {\bibinfo  {journal} {Science}\ }\textbf {\bibinfo {volume} {367}}
  (\bibinfo {year} {2020}),\ 10.1126/science.aav3392}\BibitemShut {NoStop}%
\bibitem [{\citenamefont {Valentini}\ \emph {et~al.}(2020)\citenamefont
  {Valentini}, \citenamefont {Pe{\~n}aranda}, \citenamefont {Hofmann},
  \citenamefont {Brauns}, \citenamefont {Hauschild}, \citenamefont {Krogstrup},
  \citenamefont {San-Jose}, \citenamefont {Prada}, \citenamefont {Aguado},\
  and\ \citenamefont {Katsaros}}]{valentini2020flux}%
  \BibitemOpen
  \bibfield  {author} {\bibinfo {author} {\bibfnamefont {M.}~\bibnamefont
  {Valentini}}, \bibinfo {author} {\bibfnamefont {F.}~\bibnamefont
  {Pe{\~n}aranda}}, \bibinfo {author} {\bibfnamefont {A.}~\bibnamefont
  {Hofmann}}, \bibinfo {author} {\bibfnamefont {M.}~\bibnamefont {Brauns}},
  \bibinfo {author} {\bibfnamefont {R.}~\bibnamefont {Hauschild}}, \bibinfo
  {author} {\bibfnamefont {P.}~\bibnamefont {Krogstrup}}, \bibinfo {author}
  {\bibfnamefont {P.}~\bibnamefont {San-Jose}}, \bibinfo {author}
  {\bibfnamefont {E.}~\bibnamefont {Prada}}, \bibinfo {author} {\bibfnamefont
  {R.}~\bibnamefont {Aguado}}, \ and\ \bibinfo {author} {\bibfnamefont
  {G.}~\bibnamefont {Katsaros}},\ }\href@noop {} {\bibfield  {journal}
  {\bibinfo  {journal} {arXiv preprint arXiv:2008.02348}\ } (\bibinfo {year}
  {2020})}\BibitemShut {NoStop}%
\bibitem [{\citenamefont {Blatter}\ \emph {et~al.}(1994)\citenamefont
  {Blatter}, \citenamefont {Feigel'man}, \citenamefont {Geshkenbein},
  \citenamefont {Larkin},\ and\ \citenamefont {Vinokur}}]{RMP_pinning}%
  \BibitemOpen
  \bibfield  {author} {\bibinfo {author} {\bibfnamefont {G.}~\bibnamefont
  {Blatter}}, \bibinfo {author} {\bibfnamefont {M.~V.}\ \bibnamefont
  {Feigel'man}}, \bibinfo {author} {\bibfnamefont {V.~B.}\ \bibnamefont
  {Geshkenbein}}, \bibinfo {author} {\bibfnamefont {A.~I.}\ \bibnamefont
  {Larkin}}, \ and\ \bibinfo {author} {\bibfnamefont {V.~M.}\ \bibnamefont
  {Vinokur}},\ }\href {\doibase 10.1103/RevModPhys.66.1125} {\bibfield
  {journal} {\bibinfo  {journal} {Rev. Mod. Phys.}\ }\textbf {\bibinfo {volume}
  {66}},\ \bibinfo {pages} {1125} (\bibinfo {year} {1994})}\BibitemShut
  {NoStop}%
\bibitem [{\citenamefont {Larkin}\ and\ \citenamefont
  {Ovchinnikov}(1979)}]{larkin1979pinning}%
  \BibitemOpen
  \bibfield  {author} {\bibinfo {author} {\bibfnamefont {A.}~\bibnamefont
  {Larkin}}\ and\ \bibinfo {author} {\bibfnamefont {Y.~N.}\ \bibnamefont
  {Ovchinnikov}},\ }\href@noop {} {\bibfield  {journal} {\bibinfo  {journal}
  {Journal of Low Temperature Physics}\ }\textbf {\bibinfo {volume} {34}},\
  \bibinfo {pages} {409} (\bibinfo {year} {1979})}\BibitemShut {NoStop}%
\bibitem [{\citenamefont {Thuneberg}\ \emph {et~al.}(1982)\citenamefont
  {Thuneberg}, \citenamefont {Kurkij\"arvi},\ and\ \citenamefont
  {Rainer}}]{pinning_Thuneberg}%
  \BibitemOpen
  \bibfield  {author} {\bibinfo {author} {\bibfnamefont {E.~V.}\ \bibnamefont
  {Thuneberg}}, \bibinfo {author} {\bibfnamefont {J.}~\bibnamefont
  {Kurkij\"arvi}}, \ and\ \bibinfo {author} {\bibfnamefont {D.}~\bibnamefont
  {Rainer}},\ }\href {\doibase 10.1103/PhysRevLett.48.1853} {\bibfield
  {journal} {\bibinfo  {journal} {Phys. Rev. Lett.}\ }\textbf {\bibinfo
  {volume} {48}},\ \bibinfo {pages} {1853} (\bibinfo {year}
  {1982})}\BibitemShut {NoStop}%
\bibitem [{\citenamefont {Klaassen}\ \emph {et~al.}(2001)\citenamefont
  {Klaassen}, \citenamefont {Doornbos}, \citenamefont {Huijbregtse},
  \citenamefont {van~der Geest}, \citenamefont {Dam},\ and\ \citenamefont
  {Griessen}}]{pinning_Klaassen}%
  \BibitemOpen
  \bibfield  {author} {\bibinfo {author} {\bibfnamefont {F.~C.}\ \bibnamefont
  {Klaassen}}, \bibinfo {author} {\bibfnamefont {G.}~\bibnamefont {Doornbos}},
  \bibinfo {author} {\bibfnamefont {J.~M.}\ \bibnamefont {Huijbregtse}},
  \bibinfo {author} {\bibfnamefont {R.~C.~F.}\ \bibnamefont {van~der Geest}},
  \bibinfo {author} {\bibfnamefont {B.}~\bibnamefont {Dam}}, \ and\ \bibinfo
  {author} {\bibfnamefont {R.}~\bibnamefont {Griessen}},\ }\href {\doibase
  10.1103/PhysRevB.64.184523} {\bibfield  {journal} {\bibinfo  {journal} {Phys.
  Rev. B}\ }\textbf {\bibinfo {volume} {64}},\ \bibinfo {pages} {184523}
  (\bibinfo {year} {2001})}\BibitemShut {NoStop}%
\bibitem [{\citenamefont {Jiang}\ \emph {et~al.}(2019)\citenamefont {Jiang},
  \citenamefont {Dai},\ and\ \citenamefont {Wang}}]{MZM_Kun}%
  \BibitemOpen
  \bibfield  {author} {\bibinfo {author} {\bibfnamefont {K.}~\bibnamefont
  {Jiang}}, \bibinfo {author} {\bibfnamefont {X.}~\bibnamefont {Dai}}, \ and\
  \bibinfo {author} {\bibfnamefont {Z.}~\bibnamefont {Wang}},\ }\href {\doibase
  10.1103/PhysRevX.9.011033} {\bibfield  {journal} {\bibinfo  {journal} {Phys.
  Rev. X}\ }\textbf {\bibinfo {volume} {9}},\ \bibinfo {pages} {011033}
  (\bibinfo {year} {2019})}\BibitemShut {NoStop}%
\bibitem [{\citenamefont {Schnyder}\ \emph {et~al.}(2008)\citenamefont
  {Schnyder}, \citenamefont {Ryu}, \citenamefont {Furusaki},\ and\
  \citenamefont {Ludwig}}]{classification1}%
  \BibitemOpen
  \bibfield  {author} {\bibinfo {author} {\bibfnamefont {A.~P.}\ \bibnamefont
  {Schnyder}}, \bibinfo {author} {\bibfnamefont {S.}~\bibnamefont {Ryu}},
  \bibinfo {author} {\bibfnamefont {A.}~\bibnamefont {Furusaki}}, \ and\
  \bibinfo {author} {\bibfnamefont {A.~W.~W.}\ \bibnamefont {Ludwig}},\ }\href
  {\doibase 10.1103/PhysRevB.78.195125} {\bibfield  {journal} {\bibinfo
  {journal} {Phys. Rev. B}\ }\textbf {\bibinfo {volume} {78}},\ \bibinfo
  {pages} {195125} (\bibinfo {year} {2008})}\BibitemShut {NoStop}%
\bibitem [{\citenamefont {Ryu}\ \emph {et~al.}(2010)\citenamefont {Ryu},
  \citenamefont {Schnyder}, \citenamefont {Furusaki},\ and\ \citenamefont
  {Ludwig}}]{classification2}%
  \BibitemOpen
  \bibfield  {author} {\bibinfo {author} {\bibfnamefont {S.}~\bibnamefont
  {Ryu}}, \bibinfo {author} {\bibfnamefont {A.~P.}\ \bibnamefont {Schnyder}},
  \bibinfo {author} {\bibfnamefont {A.}~\bibnamefont {Furusaki}}, \ and\
  \bibinfo {author} {\bibfnamefont {A.~W.~W.}\ \bibnamefont {Ludwig}},\ }\href
  {\doibase 10.1088/1367-2630/12/6/065010} {\bibfield  {journal} {\bibinfo
  {journal} {New Journal of Physics}\ }\textbf {\bibinfo {volume} {12}},\
  \bibinfo {pages} {065010} (\bibinfo {year} {2010})}\BibitemShut {NoStop}%
\bibitem [{\citenamefont {Kobayashi}\ and\ \citenamefont
  {Sato}(2015)}]{rotation_Sato}%
  \BibitemOpen
  \bibfield  {author} {\bibinfo {author} {\bibfnamefont {S.}~\bibnamefont
  {Kobayashi}}\ and\ \bibinfo {author} {\bibfnamefont {M.}~\bibnamefont
  {Sato}},\ }\href {\doibase 10.1103/PhysRevLett.115.187001} {\bibfield
  {journal} {\bibinfo  {journal} {Phys. Rev. Lett.}\ }\textbf {\bibinfo
  {volume} {115}},\ \bibinfo {pages} {187001} (\bibinfo {year}
  {2015})}\BibitemShut {NoStop}%
\bibitem [{\citenamefont {Fang}\ \emph {et~al.}(2017)\citenamefont {Fang},
  \citenamefont {Bernevig},\ and\ \citenamefont {Gilbert}}]{rotation_CFang}%
  \BibitemOpen
  \bibfield  {author} {\bibinfo {author} {\bibfnamefont {C.}~\bibnamefont
  {Fang}}, \bibinfo {author} {\bibfnamefont {B.~A.}\ \bibnamefont {Bernevig}},
  \ and\ \bibinfo {author} {\bibfnamefont {M.~J.}\ \bibnamefont {Gilbert}},\
  }\href@noop {} {\bibfield  {journal} {\bibinfo  {journal} {arXiv preprint
  arXiv:1701.01944}\ } (\bibinfo {year} {2017})}\BibitemShut {NoStop}%
\bibitem [{\citenamefont {Zhang}\ and\ \citenamefont
  {Liu}(2018)}]{rotation_RXZhang}%
  \BibitemOpen
  \bibfield  {author} {\bibinfo {author} {\bibfnamefont {R.-X.}\ \bibnamefont
  {Zhang}}\ and\ \bibinfo {author} {\bibfnamefont {C.-X.}\ \bibnamefont
  {Liu}},\ }\href {\doibase 10.1103/PhysRevLett.120.156802} {\bibfield
  {journal} {\bibinfo  {journal} {Phys. Rev. Lett.}\ }\textbf {\bibinfo
  {volume} {120}},\ \bibinfo {pages} {156802} (\bibinfo {year}
  {2018})}\BibitemShut {NoStop}%
\bibitem [{\citenamefont {Zhang}\ \emph {et~al.}(2020)\citenamefont {Zhang},
  \citenamefont {Hsu},\ and\ \citenamefont {Das~Sarma}}]{rotation_HOTDS}%
  \BibitemOpen
  \bibfield  {author} {\bibinfo {author} {\bibfnamefont {R.-X.}\ \bibnamefont
  {Zhang}}, \bibinfo {author} {\bibfnamefont {Y.-T.}\ \bibnamefont {Hsu}}, \
  and\ \bibinfo {author} {\bibfnamefont {S.}~\bibnamefont {Das~Sarma}},\ }\href
  {\doibase 10.1103/PhysRevB.102.094503} {\bibfield  {journal} {\bibinfo
  {journal} {Phys. Rev. B}\ }\textbf {\bibinfo {volume} {102}},\ \bibinfo
  {pages} {094503} (\bibinfo {year} {2020})}\BibitemShut {NoStop}%
\bibitem [{\citenamefont {Kopasov}\ and\ \citenamefont
  {Mel'nikov}(2020)}]{rotation_fluxshell}%
  \BibitemOpen
  \bibfield  {author} {\bibinfo {author} {\bibfnamefont {A.~A.}\ \bibnamefont
  {Kopasov}}\ and\ \bibinfo {author} {\bibfnamefont {A.~S.}\ \bibnamefont
  {Mel'nikov}},\ }\href {\doibase 10.1103/PhysRevB.101.054515} {\bibfield
  {journal} {\bibinfo  {journal} {Phys. Rev. B}\ }\textbf {\bibinfo {volume}
  {101}},\ \bibinfo {pages} {054515} (\bibinfo {year} {2020})}\BibitemShut
  {NoStop}%
\bibitem [{\citenamefont {Kitaev}(2001)}]{Kitaev_2001}%
  \BibitemOpen
  \bibfield  {author} {\bibinfo {author} {\bibfnamefont {A.~Y.}\ \bibnamefont
  {Kitaev}},\ }\href {\doibase 10.1070/1063-7869/44/10s/s29} {\bibfield
  {journal} {\bibinfo  {journal} {Physics-Uspekhi}\ }\textbf {\bibinfo {volume}
  {44}},\ \bibinfo {pages} {131} (\bibinfo {year} {2001})}\BibitemShut
  {NoStop}%
\bibitem [{\citenamefont {Machida}\ \emph {et~al.}(2019)\citenamefont
  {Machida}, \citenamefont {Sun}, \citenamefont {Pyon}, \citenamefont {Takeda},
  \citenamefont {Kohsaka}, \citenamefont {Hanaguri}, \citenamefont {Sasagawa},\
  and\ \citenamefont {Tamegai}}]{machida2019zero}%
  \BibitemOpen
  \bibfield  {author} {\bibinfo {author} {\bibfnamefont {T.}~\bibnamefont
  {Machida}}, \bibinfo {author} {\bibfnamefont {Y.}~\bibnamefont {Sun}},
  \bibinfo {author} {\bibfnamefont {S.}~\bibnamefont {Pyon}}, \bibinfo {author}
  {\bibfnamefont {S.}~\bibnamefont {Takeda}}, \bibinfo {author} {\bibfnamefont
  {Y.}~\bibnamefont {Kohsaka}}, \bibinfo {author} {\bibfnamefont
  {T.}~\bibnamefont {Hanaguri}}, \bibinfo {author} {\bibfnamefont
  {T.}~\bibnamefont {Sasagawa}}, \ and\ \bibinfo {author} {\bibfnamefont
  {T.}~\bibnamefont {Tamegai}},\ }\href {\doibase 10.1038/s41563-019-0397-1}
  {\bibfield  {journal} {\bibinfo  {journal} {Nature materials}\ }\textbf
  {\bibinfo {volume} {18}},\ \bibinfo {pages} {811} (\bibinfo {year}
  {2019})}\BibitemShut {NoStop}%
\bibitem [{\citenamefont {Kong}\ \emph {et~al.}(2020)\citenamefont {Kong},
  \citenamefont {Cao}, \citenamefont {Zhu}, \citenamefont {Papaj},
  \citenamefont {Dai}, \citenamefont {Li}, \citenamefont {Fan}, \citenamefont
  {Liu}, \citenamefont {Yang}, \citenamefont {Wang} \emph
  {et~al.}}]{kong2020tunable}%
  \BibitemOpen
  \bibfield  {author} {\bibinfo {author} {\bibfnamefont {L.}~\bibnamefont
  {Kong}}, \bibinfo {author} {\bibfnamefont {L.}~\bibnamefont {Cao}}, \bibinfo
  {author} {\bibfnamefont {S.}~\bibnamefont {Zhu}}, \bibinfo {author}
  {\bibfnamefont {M.}~\bibnamefont {Papaj}}, \bibinfo {author} {\bibfnamefont
  {G.}~\bibnamefont {Dai}}, \bibinfo {author} {\bibfnamefont {G.}~\bibnamefont
  {Li}}, \bibinfo {author} {\bibfnamefont {P.}~\bibnamefont {Fan}}, \bibinfo
  {author} {\bibfnamefont {W.}~\bibnamefont {Liu}}, \bibinfo {author}
  {\bibfnamefont {F.}~\bibnamefont {Yang}}, \bibinfo {author} {\bibfnamefont
  {X.}~\bibnamefont {Wang}},  \emph {et~al.},\ }\href@noop {} {\bibfield
  {journal} {\bibinfo  {journal} {arXiv preprint arXiv:2010.04735}\ } (\bibinfo
  {year} {2020})}\BibitemShut {NoStop}%
\end{thebibliography}%

%\begin{thebibliography}{vortex_Majorana}
%
%
%\end{thebibliography}

%\end{bibunit}

\appendix

\begin{widetext}

\section{Local SOC induced by impurity}

We take the 2D SC for instance to illustrate the effective SOC induced by the impurity. If a impurity is introduced in a homogeneous SC, additional to the on-site scattering potential it will break the inversion symmetry of the system locally. In the presence of rotational symmetry, such symmetry-breaking effect can induce an effective electric field in the radial direction, as sketched in Fig.\ref{sketch_SOC}. This leads to a Rashba SOC with the following form
\begin{eqnarray}\label{impurity_SOC1}
\mathcal{H}^{\text{im}}_{\text{soc}} = \alpha(r) (\hat{{\bf r}} \times {\bf k} ) \cdot {\bf \sigma}.
\end{eqnarray}
where $\hat{{\bf r}} = (\cos\theta, \sin\theta)$ is the direction of the effective electric field. Since it arises from the local symmetry-breaking effect, the strength of the SOC $\alpha(r)$ decays as it goes away from the impurity.

The impurity induced Rashba SOC depicted above can be generalized to the 3D SC text by replacing $\hat{{\bf r}}$ with $\hat{{\bf r}}_\perp = (\cos\theta_\perp, \sin\theta_\perp, 0)$. After some algebra, we can get the specific form of the SOC in the main text
\begin{eqnarray}\label{impurity_SOC2}
\mathcal{H}^{\text{im}}_{\text{soc}} &=& \alpha(r) [ k_z (\sin\theta_\perp\sigma_x-\cos\theta_\perp\sigma_y) + (k_y\cos\theta_\perp-k_x\sin\theta_\perp)\sigma_z ] \nonumber\\
&=& i\alpha(r) k_z ( e^{-i\theta_\perp} \sigma_+ - e^{i\theta_\perp} \sigma_- ) - \frac{\alpha(r)}{r} \hat{l}_z \sigma_z,
\end{eqnarray}
where $\sigma_\pm = (\sigma_x \pm i\sigma_y)/2$ and $\hat{l}_z = -i\partial_{\theta_\perp}$.

\begin{figure}
\centerline{\includegraphics[width=0.4\textwidth]{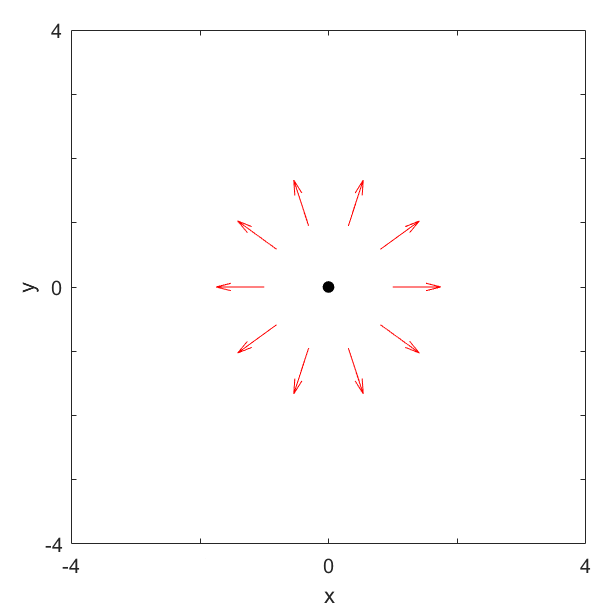}}
\caption{(color online) Sketch of the impurity induced local SOC. The impurity is represented by the black dot, and the effective electric field induced by the impurity is illustrated by the red arrow lines.
\label{sketch_SOC}}
\end{figure}

%
%\section{Effect of SOC on the VBSs}
%
%In this part, we treat the SOC induced by the impurity perturbatively and show its influence on the VBSs in the 2D superconductor in the main text, which is also the $k_z = 0$ condition of the 3D superconductor in the main text. As mentioned, the wave functions of the VBSs take the form $| \varphi_{L_z}({\bf r}) \rangle = e^{iL_z\theta} ( \chi_1(r), e^{i\theta}\chi_2(r), e^{-i\theta}\chi_3(r), \chi_4(r) )^T$, with $L_z$ the angular momentum. The SOC induced by the impurity contributes the following energy correction to the VBSs
%\begin{eqnarray}\label{SOC_perturbation}
%\Delta E_{\text{SOC}} = \int dr [ L_z( |\chi_2(r)|^2 + |\chi_3(r)|^2 - |\chi_1(r)|^2 - |\chi_4(r)|^2 ) + ( |\chi_2(r)|^2 - |\chi_3(r)|^2 ) ] \alpha(r).
%\end{eqnarray}

\section{Calculating VBSs based the Bessel functions}

To solve systems with continuum rotational symmetry, it is convenient to take advantage of the Bessel functions in the disc or cylinder geometry. The Hamiltonian in the main text without impurity reads as
\begin{eqnarray}\label{H_example}
\mathcal{H} &=& \int d{\bf r} \psi^\dagger({\bf r})h({\bf r})\psi({\bf r}),\\
h({\bf r}) &=& \left(
            \begin{array}{cccc}
            \frac{{\bf k}^2}{2m}-\mu     &      0      &       \Delta(r)e^{i\theta}      &     0     \\
            0      &      \frac{{\bf k}^2}{2m}-\mu     &     0     &    \Delta(r)e^{i\theta}      \\
            \Delta(r)e^{-i\theta}     &     0     &     -\frac{{\bf k}^2}{2m}+\mu     &     0     \\
            0     &    \Delta(r)e^{-i\theta}      &     0    &    -\frac{{\bf k}^2}{2m}+\mu       \\
            \end{array}
          \right), \nonumber
\end{eqnarray}
where we have used the same basis as that in the main text. As mentioned in the main text, the vortex line preserves the translational symmetry along the $z$-direction. Thus $k_z$ appears as a good quantum number, and we take periodic boundary condition in the $z$-direction by transforming $k_z \rightarrow \sin k_z$ and $k_z^2 \rightarrow 2(1-\cos k_z)$. Within the $xy$-plane, we can expand the system in the following form
\begin{eqnarray}\label{2D_expansion}
\psi({\bf r}) = \sum_{n,j}c_{n,j}\frac{1}{\sqrt{2\pi}}e^{in\theta}\tilde{J_n}(\beta^n_jr/R),
\end{eqnarray}
where $c_{n,j}$ annihilates an electron (creates a hole) with quantum number $\{n, j\}$, $R$ is the radius of the cylinder and $\tilde{J_n}(\beta^n_jr/R)$ is the $n$-th order normalized Bessel function of the first kind with $\beta^n_j$ its $j$-th zero. Notice that we have omit $k_z$ in the notation for simplicity here. In Eq.\eqref{2D_expansion} we can do the expansion, because the eigenfunctions of the rotation operator and the Bessel functions constitute an orthogonal complete basis in the 2D space
\begin{eqnarray}\label{orthogonal}
\frac{1}{2\pi}\int_0^{2\pi} d\theta e^{-im\theta}e^{in\theta} = \delta_{mn}, \nonumber\\
\int_0^R dr \tilde{J_n}(\beta^n_jr/R) \tilde{J_n}(\beta^n_lr/R) = \delta_{jl}.
\end{eqnarray}
The most important advantage of this choice is that, if we solve the system in a finite-size cylinder geometry the boundary condition at $r = R$, $i.e.$ the wave function vanishes at $r = R$, can be fulfilled automatically. In the numerical calculations, we also need to do a cut-off on the zeros of the Bessel functions $\beta^n_j$, which is reasonable because large $\beta^n_j$ corresponds to the fast-oscillating high-energy modes.

With the expansion in Eq.\eqref{2D_expansion}, the whole system is decoupled into different rotational invariant channels
\begin{eqnarray}\label{2D_decouple}
\mathcal{H} &=& \oplus \mathcal{H}_n,
\end{eqnarray}
where $n$ represents the angular momentum of each channel. For each $\mathcal{H}_n$, its basis is $\psi_n = (c_{n,j,\uparrow}, c_{n+1,j,\downarrow}, c_{n-1,j,\downarrow}^\dag, -c_{n,j,\uparrow}^\dag)^T$, with no constraint on $j$. In fact, the system in Eq.\eqref{H_example} can be further decoupled according to spin. However, if the SOC term in Eq.\eqref{impurity_SOC2} is taken into consideration, the two spin subspaces will hybridize and the whole system can only be decoupled according to the angular momentum. If we cut off the zeros of the Bessel functions at $\beta^N_j$, each $\mathcal{H}_n$ is a $4N \times 4N$ matrix and the matrix elements are
\begin{eqnarray}\label{matrix_elements}
\mathcal{H}_{n,jl}^{11} &=& (\frac{1}{2m}(\frac{\beta^n_j}{R})^2-\mu)\delta_{jl} + n \int_0^R d\rho \alpha(r) J_n(\beta^n_j\rho/R) J_{n}(\beta^{n}_l\rho/R) \frac{\sqrt{2}}{RJ_{n+1}(\beta^n_j)} \frac{\sqrt{2}}{RJ_{n+1}(\beta^{n}_l)}, \nonumber\\
\mathcal{H}_{n,jl}^{12} &=& ik_z \int_0^R d\rho \rho \alpha(r) J_n(\beta^n_j\rho/R) J_{n+1}(\beta^{n+1}_l\rho/R) \frac{\sqrt{2}}{RJ_{n+1}(\beta^n_j)} \frac{\sqrt{2}}{RJ_{n+2}(\beta^{n+1}_l)}, \nonumber\\
\mathcal{H}_{n,jl}^{13} &=& \int_0^R d\rho \rho \Delta(r) J_n(\beta^n_j\rho) J_{n-1}(\beta^{n-1}_l\rho) \frac{\sqrt{2}}{RJ_{n+1}(\beta^n_j)} \frac{\sqrt{2}}{RJ_n(\beta^{n-1}_l)},
\end{eqnarray}
where we have listed several typical terms and other terms can be calculated similarly. In the numerical calculations, we set $R = 150$ and cut off the Bessel functions at their 120-th zeros.

\section{Scattering potential of magnetic and nonmagnetic impurities in vortices}

In this part, we decouple the scattering potentials of different kinds of impurities into different rotational invariant channels based on the expansion in Eq.\eqref{2D_expansion}. We start with a review of the polynomial form of the Bessel functions of the first kind, which turns out to be
\begin{eqnarray}\label{Bessel_polynomial}
J_n(x) = \sum_{k=0}^\infty \frac{(-1)^k}{k!(n+k)!} (\frac{x}{2})^{2k+n}.
\end{eqnarray}
From Eq.\eqref{Bessel_polynomial}, it is obvious that $J_0(0) = 1$ while $J_n(x) = 0$ for any $n \neq 0$. Since the impurity we consider is the $\delta$-function type, it can only affect $\mathcal{H}_n$ related to $\tilde{J_0}(\beta^n_jr/R)$, namely the $n = 0$ and $n = \pm 1$ channels.

The scattering potentials of the magnetic and nonmagnetic impurities can be described by
\begin{eqnarray}\label{potential_mag_nonmag}
\mathcal{H}_{\text{mag}} = S({\bf r}) \cdot \sigma_z \tau_0, \nonumber\\
\mathcal{H}_{\text{nonmag}} = S({\bf r}) \cdot \sigma_0 \tau_z.
\end{eqnarray}
The effects of the impurities can be decoupled into two subspaces, $\psi^\uparrow = (c_{{\bf r} \uparrow}, c_{{\bf r} \downarrow}^\dag)^T$ and $\psi^\downarrow = (c_{{\bf r} \downarrow}, c_{{\bf r} \uparrow}^\dag)^T$. In general, $\psi^\uparrow$ and $\psi^\downarrow$ are related by the particle-hole symmetry. In the two subspaces, the contribution of the impurities are
\begin{eqnarray}\label{potential_decouple}
\mathcal{H}_{\text{mag}}^\uparrow &=& S({\bf r}) \cdot \tau_0, \qquad
\mathcal{H}_{\text{mag}}^\downarrow = -S({\bf r}) \cdot \tau_0, \nonumber\\
\mathcal{H}_{\text{nonmag}}^\uparrow &=& S({\bf r}) \cdot \tau_z, \qquad
\mathcal{H}_{\text{nonmag}}^\downarrow = S({\bf r}) \cdot \tau_z.
\end{eqnarray}
For the spin-up subspace, the $\delta$-function type impurity can only affect the following two channels, $\psi^\uparrow_1 = (c_{1,j,\uparrow}, c_{0,j,\downarrow}^\dag)^T$ and $\psi^\uparrow_0 = (c_{0,j,\uparrow}, c_{-1,j,\downarrow}^\dag)^T$. In the above two subspaces, the matrix form of the Hamiltonian in Eq.\eqref{potential_decouple} can be expressed as
\begin{eqnarray}\label{impurity_decouple_up}
\mathcal{H}_{\text{mag},1}^\uparrow &=& J \left(
                                           \begin{array}{cc}
                                            0           &            0            \\
                                            0           &            h            \\
                                           \end{array}
                                         \right), \qquad
\mathcal{H}_{\text{mag},0}^\uparrow = J \left(
                                          \begin{array}{cc}
                                           h           &            0           \\
                                           0           &            0           \\
                                          \end{array}
                                        \right), \nonumber\\
\mathcal{H}_{\text{nonmag},1}^\uparrow &=& V \left(
                                              \begin{array}{cc}
                                               0           &            0            \\
                                               0           &            -h           \\
                                              \end{array}
                                            \right), \qquad
\mathcal{H}_{\text{nonmag},0}^\uparrow = V \left(
                                             \begin{array}{cc}
                                              h           &            0           \\
                                              0           &            0           \\
                                             \end{array}
                                           \right),
\end{eqnarray}
where $J$ and $V$ are the strength of the scattering potential for the magnetic and nonmagnetic impurities respectively. In Eq.\eqref{impurity_decouple_up}, $h$ is a $N \times N$ matrix where $N$ characterize the cut-off of the zeros of the Bessel functions and
\begin{eqnarray}\label{impurity_matrix_element}
h_{ij} &=& \int_0^R d\rho \delta(\rho) \tilde{J_0}(\beta^0_i\rho/R) \tilde{J_0}(\beta^0_j\rho/R) \nonumber\\
&=& \tilde{J_0}(0) \tilde{J_0}(0).
\end{eqnarray}
Similarly, in the spin-down subspace we consider two the channels $\psi^\downarrow_{-1} = (c_{0,j,\downarrow}, -c_{-1,j,\uparrow}^\dag)^T$ and $\psi^\downarrow_0 = (c_{1,j,\downarrow}, -c_{0,j,\uparrow}^\dag)^T$, where we have
\begin{eqnarray}\label{impurity_decouple_down}
\mathcal{H}_{\text{mag},-1}^\downarrow &=& J \left(
                                             \begin{array}{cc}
                                              -h          &            0            \\
                                              0           &            0            \\
                                             \end{array}
                                           \right), \qquad
\mathcal{H}_{\text{mag},0}^\downarrow = J \left(
                                            \begin{array}{cc}
                                             0           &            0            \\
                                             0           &            -h           \\
                                            \end{array}
                                          \right), \nonumber\\
\mathcal{H}_{\text{nonmag},-1}^\downarrow &=& V \left(
                                                \begin{array}{cc}
                                                 h           &            0            \\
                                                 0           &            0            \\
                                                \end{array}
                                              \right), \qquad
\mathcal{H}_{\text{nonmag},0}^\downarrow = V \left(
                                               \begin{array}{cc}
                                                0           &            0           \\
                                                0           &            -h          \\
                                               \end{array}
                                             \right).
\end{eqnarray}

In the absence of SOC, the contributions of the Hamiltonian in Eq.\eqref{H_example} are all the same for the spin-up and spin-down subspaces. Therefore, based on the results in Eq.\eqref{impurity_decouple_up} and Eq.\eqref{impurity_decouple_down} we can come to the conclusions in the main text: In the $n = \pm 1$ channel, a magnetic impurity is equivalent to a nonmagnetic impurity with opposite scattering strength; while in the $n = 0$ subspace, a magnetic impurity is equivalent to a nonmagnetic impurity with the same scattering strength.

The above results can be understood by doing a gauge transformation: $\tilde{c}_{{\bf r} \sigma}^\dagger = c_{{\bf r} \sigma}^\dagger e^{i\theta/2}$, $\tilde{c}_{{\bf r} \sigma} = c_{{\bf r} \sigma} e^{-i\theta/2}$. This transformation makes the Cooper pair carry no angular momentum, but the electron part and the hole part in the BdG Hamiltonian carry nonzero and opposite angular momenta. It is worth mentioning that though we focus on the VBSs problem here, the above analysis can also be applied to the chiral SCs since the Cooper pairs in a chiral SC carry nonzero angular momentum.

%In other words, the nonzero angular momentum carried by the Cooper pairs makes a impurity, no matter it is magnetic or nonmagnetic, equivalent to such a kind of impurity which is a superposition of the magnetic impurity and the nonmagnetic impurity in a conventional SC, if we compare the expressions in Eq.\ref{impurity_decouple_up} and Eq.\ref{impurity_decouple_down} with that in a homogeneous conventional SC.

\section{Weak impurities as perturbation}

In the SC considered in the main text, there are three independent parameters: the effective mass $m$, the chemical potential $\mu$ and the superconducting order parameter $\Delta_0$. As well known, in such a conventional SC the VBSs have a minimal gap about $\Delta_0^2/\mu$ which is usually a small value. The topological phase transitions of the VBSs induced by the impurities occur when the minimal gap of the VBSs closes. Therefore, the VBSs can be driven into the topologically nontrivial states by the weak impurities, as shown in the main text, and it is reasonable to treat the impurities as perturbations if we focus on the topological phase transitions of the VBSs. Based on the perturbation theory (up to first order), in the limit of vanishing SOC the energy modification for state $| \varphi_{L_z}({\bf r}) \rangle$ is
\begin{eqnarray}\label{impurity_perturbation_SM}
\Delta E &=& \int rd\theta dr \langle \varphi_{L_z}({\bf r}) | \frac{\delta(r-r_0)\delta(\theta-\theta_0)}{r} U \Gamma | \varphi_{L_z}({\bf r}) \rangle, \nonumber\\
&=& \chi_i^\ast(r_0) \Gamma_{ij} \chi_j(r_0) U,
\end{eqnarray}
where $U$ and $\Gamma$ are the scattering strength and the scattering matrix of the impurity respectively.

Eq.\eqref{impurity_perturbation_SM} indicates that, the energy modification is proportional to both the strength of the impurity and the distribution of the wave function at the impurity site. Fig.\ref{fig_distribution}(a)$\sim$(e) show the distribution of the wave function of the VBSs with the lowest excitation energy in different angular momentum channels in a free vortex. Obviously, as the angular momentum grows, the energy of the VBSs becomes larger and its wave function becomes more and more extended away from the vortex core. Therefore, in the weak impurity region only the VBSs with the lowest energy in the $L_z = 0$ and $L_z = \pm 1$ channels can be prominently affected by the impurity. We can also roughly estimate the critical impurity strength where the phase transition of the VBSs occurs based on Eq.\eqref{impurity_perturbation_SM}. Since the minimal gap of the VBSs is about $\Delta_0^2/E_F$, the phase transition of the VBSs occur when $| \Delta E | = \Delta_0^2/2E_F$. Straightforwardly, we can come to the conclusion that the critical impurity strength satisfies $J_c \propto 1/E_F$ (we take the magnetic impurity case for instance). In fig.\ref{fig_distribution}(f), we show the critical impurity strength for different chemical potentials. Apparently, it satisfies the relationship $J_c \propto 1/\mu$ well.

%the VBSs are almost impervious to the weak impurities located out of the vortex. It is worth mentioning that, in the $L_z = 0$ and $L_z = \pm 1$ channels the lowest-energy VBSs are very sensitive to the impurities located at the vortex core because their wave functions are mainly localized at the vortex and decay exponentially away from it.

\begin{figure}
\centerline{\includegraphics[width=0.8\textwidth]{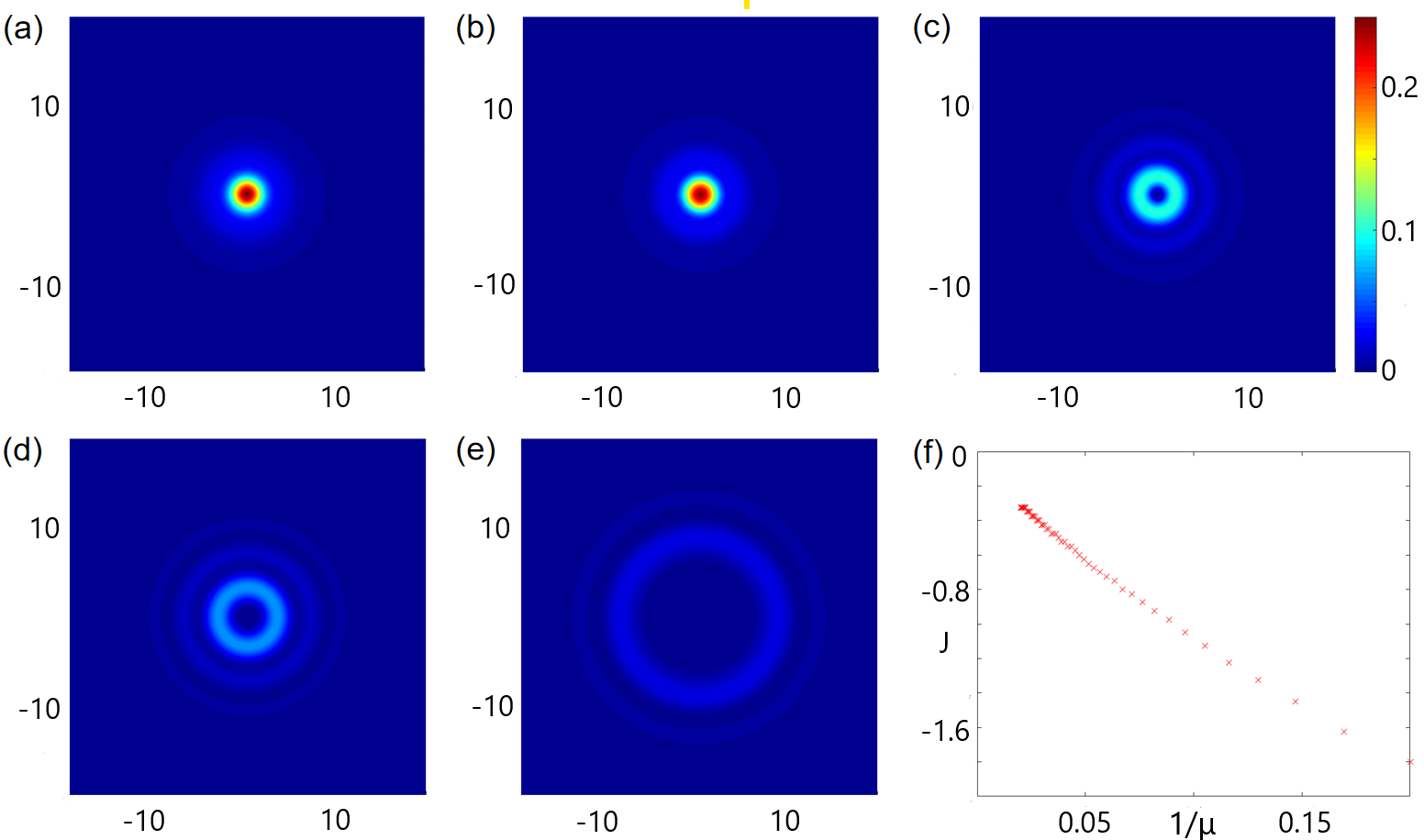}}
\caption{(color online) (a)$\sim$(e) show the distribution of the wave function of the VBSs with the lowest excitation energy in the $L_z = 0, 1, 2, 3, 8$ channels in a free vortex, and these figures share the same color bar. In (f), we take the magnetic impurity case for instance and show the relationship between the impurity strength and the chemical potential at the gap-close point of the VBSs.
\label{fig_distribution}}
\end{figure}

\begin{figure}
\centerline{\includegraphics[width=0.6\textwidth]{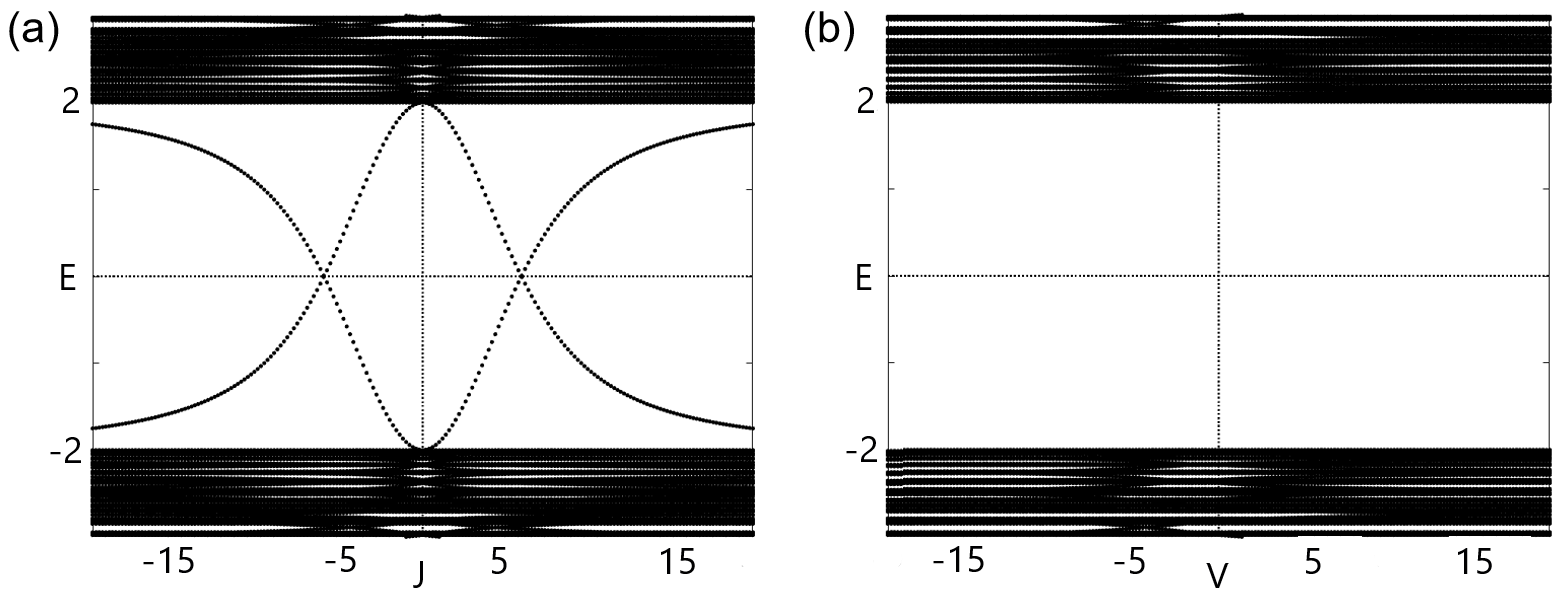}}
\caption{(color online) (a)(b) show the bound states induced by a single magnetic and nonmagnetic impurity in a 2D SC in the absence of SOC respectively.
\label{fig_YSR}}
\end{figure}

We want to emphasize that the critical impurity strength inducing the phase transition of the VBSs is rather weak, and this kind of weak impurity can not induce bound states deep in the superconducting gap if there is no vortex. We have calculated the impurity bound states in a conventional superconductor in the absence of vortices, and show the results in Fig.\ref{fig_YSR}. Obviously, no matter the impurity is magnetic or nonmagnetic (the critical impurity strength is $| J_c | = | V_c | = 1.08$ for the VBSs), it can only induce bound states near the superconducting gap at the critical impurity strength.

\section{Low-energy effective theory of the VBSs in the vortex line in a 3D SC}

In this part, based on the perturbation theory we get the effective low-energy Hamiltonian for the VBSs in the vortex line with a impurity chain (we take the nonmagnetic impurity case for instance) at its core in a 3D conventional SC. The effects of the impurity $\mathcal{H}_{\text{im}}$ includes two parts: the scattering potential in Eq.\eqref{potential_mag_nonmag} and the SOC in Eq.\eqref{impurity_SOC2}. Moreover, as mentioned in the last part only the VBSs with the lowest energy in the $L_z = 0$ and $L_z = \pm 1$ channels can be prominently affected by the impurities in the weak impurity region. Therefore, we can merely consider the two subspaces spanned by the lowest-energy states in the $\mathcal{H}_0$ and $\mathcal{H}_{1} \oplus \mathcal{H}_{-1}$ channels respectively.

In the $\mathcal{H}_0$ channel, the subspace is spanned by the two states $| S_1 \rangle = | \varphi_{0}^{\text{min}}({\bf r}) \rangle$ and $| S_2 \rangle = P | \varphi_{0}^{\text{min}}({\bf r}) \rangle$, with $P = \sigma_y \tau_y K$ the particle-hole operator. Considering that the eigenfunction of the VBS has the form $| \varphi_{0}^{\text{min}}({\bf r}) \rangle = e^{ik_z z} e^{i( - \frac{1}{2}\sigma_z + \frac{1}{2}\tau_z )\theta} (a_i\tilde{J}_0(\beta^0_i r/R), 0, b_j\tilde{J}_{-1}(\beta^{-1}_j r/R), 0)^T = e^{ik_z z} e^{i( - \frac{1}{2}\sigma_z + \frac{1}{2}\tau_z )\theta} (\chi^0_{e,\uparrow}(r), 0, \chi^0_{h,\downarrow}(r), 0)^T$, we can calculate the low-energy effective Hamiltonian as
\begin{eqnarray}\label{Lzero_effective}
\langle S_1 | \mathcal{H}_{\text{im}} | S_1 \rangle &=& -\langle S_2 | \mathcal{H}_{\text{im}} | S_2 \rangle = |E^{\text{min}}_0| - |\Delta E_0|, \nonumber \\
\langle S_1 | \mathcal{H}_{\text{im}} | S_2 \rangle &=& \langle S_2 | \mathcal{H}_{\text{im}} | S_1 \rangle^\ast = \lambda,
\end{eqnarray}
with
\begin{eqnarray}\label{Lzero_effective}
     \Delta E_0 &=& \int rd\theta dr \langle \varphi_{0}^{\text{min}}({\bf r}) | \frac{\delta(r)}{2\pi r} V \sigma_0 \tau_z | \varphi_{0}^{\text{min}}({\bf r}) \rangle - \int rd\theta dr \langle \varphi_{0}^{\text{min}}({\bf r}) | \frac{\alpha(r)}{r} \hat{l}_z \sigma_z \tau_z | \varphi_{0}^{\text{min}}({\bf r}) \rangle, \nonumber \\
&=& V \sum_{i,j} a_i^\ast a_j \tilde{J}_0(0)\tilde{J}_0(0) - 2\pi \sum_{i,j} b_i^\ast b_j \int dr \alpha(r) \tilde{J}_{-1}(\beta^{-1}_i r/R)\tilde{J}_{-1}(\beta^{-1}_j r/R), \nonumber \\
&=& V | \chi^0_{e,\uparrow}(0) |^2 - 2\pi \int dr \alpha(r) | \chi^0_{h,\downarrow}(r) |^2, \nonumber \\
     \lambda &=& \int rd\theta dr \langle \varphi_{0}^{\text{min}}({\bf r}) | i\alpha(r) k_z ( e^{-i\theta} \sigma_+ - e^{i\theta} \sigma_- ) \tau_z P| \varphi_{0}^{\text{min}}({\bf r}) \rangle, \nonumber \\
&=& 2\pi ik_z \int dr r\alpha(r) \chi^{0\ast}_{e,\uparrow}(r) \chi^{0\ast}_{h,\downarrow}(r)
 =  \lambda_0 k_z.
\end{eqnarray}
Moreover, since the Fermi surface of the system in a given $k_z$ slice becomes smaller as $k_z$ grows, $|E^{\text{min}}_0|$ becomes larger correspondingly considering that $|E^{\text{min}}_0| \approx \Delta_0^2/2\mu^\ast$ ($\mu^\ast = \mu - k_z^2/2m$ and we do not consider the $k_z$ plane where there is no Fermi surface). In summary, the low-energy theory in the $\mathcal{H}_0$ channel can be written as
\begin{eqnarray}\label{Hzero_effective}
h^0_{\text{eff}} &=& (|E^{\text{min}}_0(k_z)| - |\Delta E_0|)s_z + k_z(s_x \text{Re} \lambda_0 - s_y \text{Im} \lambda_0), \nonumber\\
&=& E_0(k_z)s_z + k_z(s_x \text{Re} \lambda_0 - s_y \text{Im} \lambda_0),
\end{eqnarray}
where $s_i$ are the Pauli matrices on behalf of the space spanned by $| S_1 \rangle$ and $| S_2 \rangle$. Obviously, Eq.\eqref{Hzero_effective} is equivalent to the Kitaev chain model.

In the $\mathcal{H}_{1} \oplus \mathcal{H}_{-1}$ channel we consider the subspace spanned by $| S_1 \rangle = | \varphi_{1}^{\text{min}}({\bf r}) \rangle$ and $| S_2 \rangle = | \varphi_{-1}^{\text{min}}({\bf r}) \rangle = P | \varphi_{1}^{\text{min}}({\bf r}) \rangle$, with $| \varphi_{1}^{\text{min}}({\bf r}) \rangle = e^{ik_z z} e^{i( 1 - \frac{1}{2}\sigma_z + \frac{1}{2}\tau_z )\theta} (a_i\tilde{J}_1(\beta^1_i r/R), 0, b_j\tilde{J}_{0}(\beta^{0}_j r/R), 0)^T = e^{ik_z z} e^{i( 1 - \frac{1}{2}\sigma_z + \frac{1}{2}\tau_z )\theta} (\chi^1_{e,\uparrow}(r), 0, \chi^1_{h,\downarrow}(r), 0)^T$. Similarly, we calculate the perturbation and summarize as follows
\begin{eqnarray}\label{Lone_effective}
\langle S_1 | \mathcal{H}_{\text{im}} | S_1 \rangle &=& -\langle S_2 | \mathcal{H}_{\text{im}} | S_2 \rangle = |E^{\text{min}}_1| - |\Delta E_1|
= -V | \chi^1_{h,\downarrow}(0) |^2 - 2\pi \int dr \alpha(r) | \chi^1_{e,\uparrow}(r) |^2, \nonumber \\
\langle S_1 | \mathcal{H}_{\text{im}} | S_2 \rangle &=& \langle S_2 | \mathcal{H}_{\text{im}} | S_1 \rangle^\ast = 0.
\end{eqnarray}
Therefore the low-energy effective model in this channel is
\begin{eqnarray}\label{Hone_effective}
h^{\pm 1}_{\text{eff}} = (|E^{\text{min}}_1(k_z)| - |\Delta E_1|)s_z.
\end{eqnarray}

\section{Topological phase transitions of VBSs induced by magnetic impurities}

In the main text, we show the gap-close-reopen processes, $i.e.$ the topological phase transitions, of the VBSs induced by the nonmagnetic impurities in a 3D SC. In this part, we show the results for the magnetic impurity case in Fig.\ref{fig_mag}. Obviously, the magnetic impurities play a similar role with the nonmagnetic impurities, as analyzed in the main text.

\begin{figure}
\centerline{\includegraphics[width=0.8\textwidth]{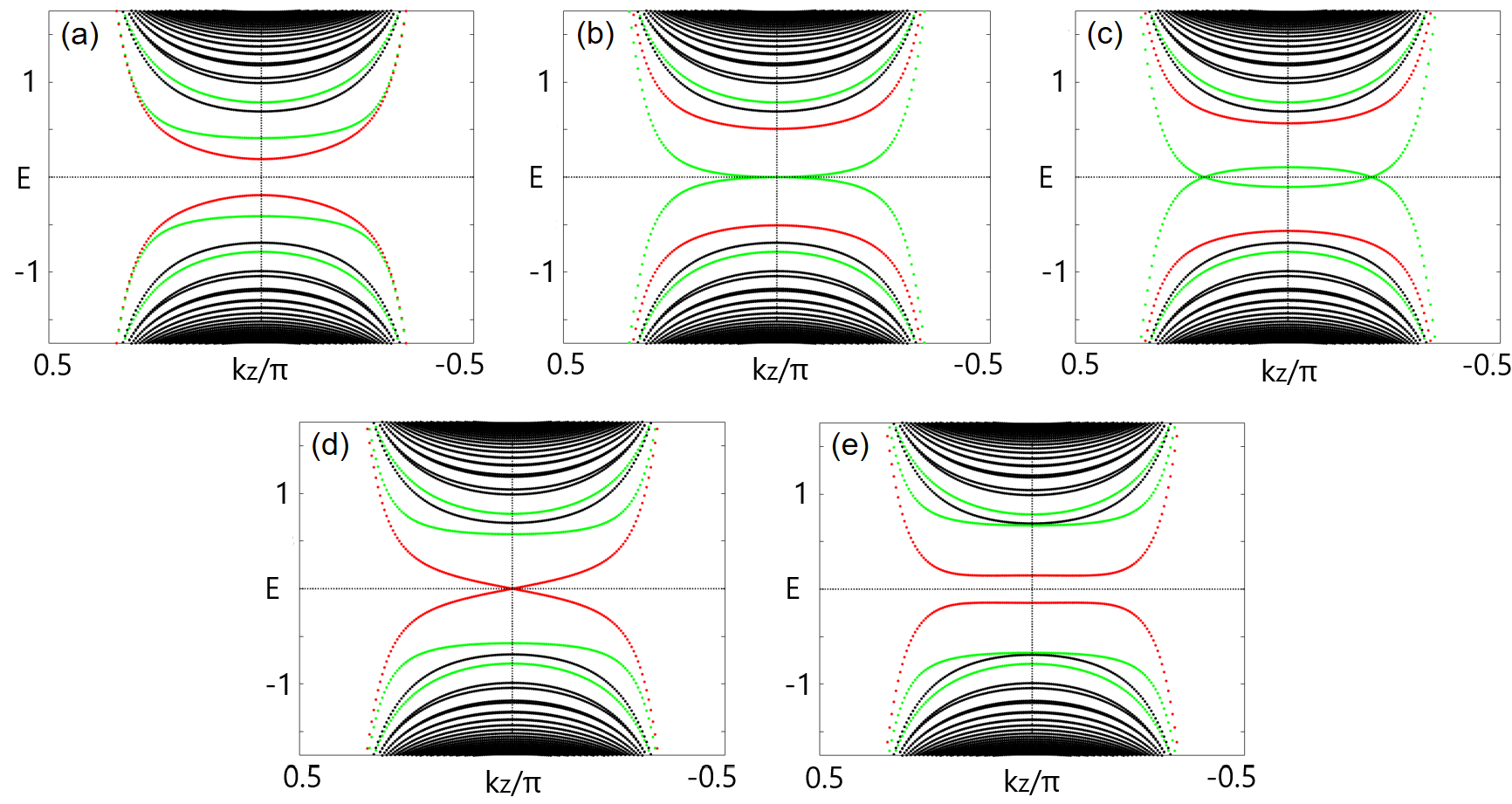}}
\caption{(color online) (a)$\sim$(e) show the energy spectrum of the VBSs in the condition of a magnetic impurity chain at the core of the vortex line, as a function of $k_z$ corresponding to impurity strength $J = 0.0$, $J = 1.47$, $J = 1.80$, $J = -0.71$ and $J = -1.20$ respectively. In the figures, we label the low-energy states in the angular momentum channel $n = 0$ with red color, while the $n = \pm 1$ channel with green color. The other parameters are the same as that in the main text.
\label{fig_mag}}
\end{figure}

\end{widetext}

\end{document}